\documentclass[12pt,a4paper]{article}

\usepackage{epsfig}
\usepackage{amsmath,amsfonts,amssymb}
\usepackage{t1enc}
\usepackage{cite}
\usepackage{colortbl}
\usepackage{pifont}

\providecommand{\openone}{\leavevmode\hbox{\small1\kern-3.8pt\normalsize1}}

\newcommand{\RE}{\text{Re}\,}
\newcommand{\IM}{\text{Im}\,}

\newcommand{\gM}{\gamma^\mu}
\newcommand{\gm}{\gamma_\mu}
\newcommand{\la}{\lambda^a}
\newcommand{\tI}{\tau^I}
\newcommand{\oh}{\textstyle \frac{1}{2}}
\newcommand{\ot}{\textstyle \frac{1}{3}}

\newcommand{\twt}{\textstyle \frac{2}{3}}
\newcommand{\ft}{\textstyle \frac{4}{3}}
\newcommand{\et}{\textstyle \frac{8}{3}}
\newcommand{\osx}{\textstyle \frac{1}{6}}

\newcommand{\os}{\textstyle \frac{1}{6}}

\newcommand{\aqq}{\alpha_{qq}}
\newcommand{\aqqp}{\alpha_{qq'}}
\newcommand{\alq}{\alpha_{\ell q}}
\newcommand{\alqp}{\alpha_{\ell q'}}

\newcommand{\auu}{\alpha_{uu}}
\newcommand{\aud}{\alpha_{ud}}
\newcommand{\audp}{\alpha_{ud'}}
\newcommand{\aeu}{\alpha_{eu}}

\newcommand{\aqu}{\alpha_{qu}}
\newcommand{\aqup}{\alpha_{qu'}}
\newcommand{\aqd}{\alpha_{qd}}
\newcommand{\aqdp}{\alpha_{qd'}}
\newcommand{\aqde}{\alpha_{qde}}
\newcommand{\alu}{\alpha_{\ell u}}
\newcommand{\aqe}{\alpha_{qe}}

\newcommand{\aqqe}{\alpha_{qq\epsilon}}
\newcommand{\aqqep}{\alpha_{qq\epsilon'}}
\newcommand{\aqle}{\alpha_{q\ell\epsilon}}
\newcommand{\alqe}{\alpha_{\ell q\epsilon}}

\newcommand{\Cqq}{C_{qq}}
\newcommand{\Cqqp}{C_{qq'}}
\newcommand{\Cuu}{C_{uu}}
\newcommand{\Cqu}{C_{qu}}
\newcommand{\Cqup}{C_{qu'}}
\newcommand{\Cqd}{C_{qd}}
\newcommand{\Cqdp}{C_{qd'}}
\newcommand{\Cud}{C_{ud}}
\newcommand{\Cudp}{C_{ud'}}
\newcommand{\Cqqe}{C_{qq\epsilon}}
\newcommand{\Cqqep}{C_{qq\epsilon'}}

\newcommand{\Clq}{C_{\ell q}}
\newcommand{\Clqp}{C_{\ell q'}}
\newcommand{\Clu}{C_{\ell u}}
\newcommand{\Ceu}{C_{eu}}
\newcommand{\Cqe}{C_{qe}}
\newcommand{\Cqde}{C_{qde}}
\newcommand{\Cqle}{C_{q \ell \epsilon}}
\newcommand{\Clqe}{C_{\ell q \epsilon}}

\newcommand{\pr}{\Pi}
\newcommand{\Om}{\Omega}
\newcommand{\vll}{\mathcal{V}_{LL}}
\newcommand{\vlr}{\mathcal{V}_{LR}}
\newcommand{\vrl}{\mathcal{V}_{RL}}
\newcommand{\vrr}{\mathcal{V}_{RR}}

\newcommand{\gris}{\cellcolor[gray]{0.8}}
\newcommand{\ok}{\checkmark}

\begin{document}

\begin{center}
\begin{Large}
{\bf Effective four-fermion operators in top physics: \\[2mm]
a roadmap}
\end{Large}

\vspace{0.5cm}
J. A. Aguilar--Saavedra \\[0.2cm] 
{\it Departamento de F\'{\i}sica Te\'orica y del Cosmos and CAFPE, \\
Universidad de Granada, E-18071 Granada, Spain}
\end{center}

\begin{abstract}
We write down a minimal basis for dimension-six gauge-invariant four-fermion operators, with some operator replacements with respect to previous ones which make it simpler for calculations. Using this basis we classify all four-fermion operator contributions involving one or two top quarks. Taking into account the different fermion chiralities, possible colour contractions and independent flavour combinations, a total number of 572 gauge-invariant operators are involved. We apply this to calculate all three-body top decay widths $t \to d_k u_i \bar d_j$,
$t \to d_k e_i^+ \nu_j$, $t \to u_k u_i \bar u_j$, $t \to u_k e_j^+ e_i^-$, $t \to u_k \bar \nu_j \nu_i$ (with $i,j,k$ generation indices) mediated by dimension-six four-fermion operators, including the interference with the Standard Model amplitudes when present. 
All single top production cross sections in $pp$, $p \bar p$ and $e^+ e^-$ collisions are calculated as well, namely $u_i d_k \to d_j t$, $\bar d_j d_k \to \bar u_i t$, $u_i \bar d_j \to \bar d_k t$, $u_i u_k \to u_j t$, $u_i \bar u_j \to \bar u_k t$, $e^+ e^- \to \bar u_k t$ and the charge conjugate processes. We also compute all top pair production cross sections, $ \bar u_i u_j \to t \bar t$, $ \bar d_i d_j \to t \bar t$, $u_i u_j \to t t$ and $e^+ e^- \to t \bar t$. Our results are completely general, without assuming any particular relation among effective operator coefficients.
\end{abstract}

\section{Introduction}

Indirect searches for physics at scales not directly accessible have proved to be fruitful in the past as, for instance, the successful prediction of the top quark mass from radiative corrections has shown. Above the electroweak symmetry breaking scale, new physics not directly observed can be probed by parameterising its effects in terms of an effective Lagrangian involving only the Standard Model (SM) fields and invariant under the  SM gauge symmetry $\text{SU}(3)_c \times \text{SU}(2)_L \times \text{U}(1)_Y$~\cite{Burges:1983zg,Leung:1984ni,Buchmuller:1985jz},
\begin{equation}
\mathcal{L}_\text{eff} = \sum \frac{C_x}{\Lambda^2} O_x + \dots \,,
\label{ec:effL}
\end{equation}
where $O_x$ are dimension-six gauge-invariant operators, $\Lambda$ is the new physics scale and $C_x$ are complex constants. Effects of dimension-eight and higher-order operators are suppressed by at least $1/\Lambda^4$, and are ignored in this work. Dimension-six gauge-invariant operators were classified in Refs.~\cite{Buchmuller:1985jz,Arzt:1994gp}, totalling 81 operators up to (many combinations of) flavour indices. Later, several of these operators have been found to be redundant~\cite{Grzadkowski:2003tf,AguilarSaavedra:2008zc} and the original list has been significantly reduced.

For top physics, the most relevant dimension-six operators are (i) those yielding top tri-linear couplings with a $W$, $Z$, photon, gluon or Higgs boson; (ii) four-fermion ones. The former have been classified in Refs.~\cite{AguilarSaavedra:2008zc} and a minimal set of top anomalous couplings has been obtained by dropping redundant operators. For the latter, the aim of this paper is precisely to perform such a classification, concentrating on the operators involving one or two top quarks. These operators can mediate top three-body decays, single top production in association with a light quark and top pair production. They will therefore be probed with a high precision at the Large Hadron Collider (LHC), which is expected to produce top quarks copiously. The phenomenology of top-gauge boson operators using a minimal basis has been investigated in detail in Refs.~\cite{AguilarSaavedra:2010nx,AguilarSaavedra:2010rx}. For four-fermion operators, there is yet a wide field to be explored.

We will begin our task by writing down a new, completely general, basis for dimension-six four-fermion operators. We will prove that it is equivalent to previous ones~\cite{Buchmuller:1985jz,Arzt:1994gp} with some redundant operators dropped and few operator replacements which make our basis more ``symmetric''. We will find some advantages when using it. First, amplitude calculations are more straightforward, as the colour and isospin structures at the operator level are simpler. Secondly, the results obtained for many observables of interest (as for instance cross sections and decay widths) are very simple in this basis due to its symmetry, and interferences between operators and also with SM contributions are trivial in most cases. For specific processes a different, particular operator selection may reduce further the interferences and give slightly more compact expressions but, in general, the expressions obtained in our basis are quite simple, given the large number of parameters involved. And, in any case, expressions for observables in terms of a different operator set are straightforward to obtain, as we will occassionally do in order to compare with previous literature.

After introducing our basis we will classify all four-fermion operators which give contributions to the effective Lagrangian involving one or two top quarks. Taking into account the different fermion chiralities, colour contractions and independent flavour combinations, a total number of 572 gauge-invariant operators are involved. But remarkably, all the contributions to the Lagrangian we are interested in can be neatly summarised in few tables of easy reading, which we expect will be useful for future four-fermion operator studies, at the very least for bookkeeping purposes. This classification allows to easily find out
which gauge-invariant operators generate four-fermion terms with one top quark, with two top quarks, or both, and the relations between these contributions implied by gauge symmetry.
The type of the terms generated determines the processes to which gauge-invariant operators can contribute, and in which their presence can be probed. Thus, relations between four-fermion contributions imply relations between processes in which new physics may manifest itself.

As a first application of this classification, we calculate all three-body top decay widths, single top and top pair cross sections in $pp$, $p \bar p$ and $e^+ e^-$ collisions, including dimension-six four-fermion operators, the SM contribution and their interference. They are:
\begin{itemize}
\item Charged current decays $t \to d_k u_i d_j$, $t \to d_k e_i^+ \nu_j$ and production processes $u_i d_k \to d_j t$, $\bar d_j d_k \to \bar u_i t$, $u_i \bar d_j \to \bar d_k t$ (the charge conjugate processes are also understood). They involve SM contributions,
which are very suppressed by small Cabibbo-Kobayashi-Maskawa (CKM) mixing angles $V_{3k}$ for $k=1,2$, as well as four-fermion ones.
\item Flavour-changing neutral (FCN) decays $t \to u_k u_i \bar u_j$, $t \to u_k e_j^+ e_i^-$, $t \to u_k \bar \nu_j \nu_i$ and production processes $u_i u_k \to u_j t$, $u_i \bar u_j \to \bar u_k t$, $e^+ e^- \to \bar u_k t$ which do not take place at the tree level in the SM and are suppressed at one loop by the Glashow-Iliopoulos-Maiani mechanism~\cite{Glashow:1970gm}. In this case, SM contributions can be safely neglected. (Strictly speaking, four-fermion amplitudes do not involve neutral currents, but it is still useful to use this notation for processes with four quarks of charge $2/3$.)
\item Top pair production processes: $t \bar t$ production $\bar u_i u_j \to t \bar t$,
$\bar d_i d_j \to t \bar t$, $e^+ e^- \to t \bar t$, which have a SM contribution, and like-sign top pair production $u_i u_j \to t t$ which is absent in the SM at the tree level. In particular, we give expressions to calculate the top forward-backward (FB) asymmetry at Tevatron including all contributing four-fermion operators.
\end{itemize}
The explicit expressions provided for these observables are relatively simple. Nevertheless, there are a plethora of processes studied and keeping a reasonable paper length requires some amount of compact notation, giving observables 
such as cross sections in terms of gauge-invariant operator coefficients and numerical factors, collected in tables for LHC with a centre of mass (CM) energy of 14 and 7 TeV, and for Tevatron.

It is not our aim to explore the phenomenological consequences of the results derived in this paper, although we will in some cases comment about the implications of these results. Such studies, to be properly addressed, require either to treat the independent parameters (operator coefficients) as effectively independent, or a well-based assumption regarding the relations among them. After all, a gauge-invariant operator basis is a basis in which heavy new physics contributions can be parameterised. One would not expect that any kind of new physics, when integrated out from the Lagrangian, yields effective operators with unrelated coefficients, all of the same order and without ``cancellations''. On the contrary, the opposite behaviour is often found~\cite{delAguila:2000aa,delAguila:2010mx}: heavy new physics when integrated out gives effective operators with correlated coefficients. With this philosophy, we will ignore the common prejudice which sets to zero the coefficients of operators containing terms which could affect $B$ physics, invoking the absence of cancellations between effective operator contributions. That possibility, which may appear to be a ``fine tuning'', may well be only an effect of the choice of basis. Examples are known~\cite{delAguila:2000aa,AguilarSaavedra:2010sq} for which these apparent cancellations are not only natural but {\em required} by the nature of the new heavy physics which is integrated out to yield effective operators.

A final point deserves mention here. In our calculations we keep terms linear in operator coefficients, proportional to $1/\Lambda^2$, and quadratic ones proportional to 
$1/\Lambda^4$. This is not inconsistent despite the fact that we ignore dimension-eight and higher-order operators. For processes without a SM contribution, and for fermion chiralities which do not interfere with the SM amplitudes, the lowest-order term is the $1/\Lambda^4$ one, and higher-dimension operators give contributions suppressed by higher powers of $\Lambda$. Therefore, the expansion is consistent. For fermion chiralities interfering with the SM, dimension-six operators give
linear $1/\Lambda^2$ and quadratic $1/\Lambda^4$ terms, while dimension-eight operators would give $1/\Lambda^4$ and $1/\Lambda^8$ contributions. In this case, $1/\Lambda^4$ terms are sub-leading and could be dropped, but we still keep them (there is no harm in doing that, and they can always be discarded a posteriori) as part of a complete calculation to order $1/\Lambda^4$, with some missing $1/\Lambda^4$ contributions from dimension-eight operators interfering with the SM amplitude.

The necessity to keep quadratic terms in calculations should be clear for many reasons. First, there are many new physics effects that cannot be properly addressed only with the operators interfering with the SM. Actually, operators which do not have interference with the SM are the ones mediating genuinely new physics effects, beyond corrections to SM processes. FCN processes, which are extremely suppressed within the SM, constitute one classical example but there are several other ones, as chirality-breaking effects for light quarks (see Ref.~\cite{AguilarSaavedra:2010sq} for a detailed discussion). Such effects, absent in the SM, could then be visible even if suppressed by $1/\Lambda^4$.
On the other hand, nothing guarantees that, if new heavy physics manifests itself at low energies, it can be parameterised precisely by the operators interfering with the SM.
A third reason is that, as we will find in the following, the quadratic $1/\Lambda^4$ corrections from non-interfering operators can be in some cases as large as the linear $1/\Lambda^2$ corrections from interference terms.

The rest of this paper is structured as follows. In section~\ref{sec:2} we write our four-fermion operator basis and in section~\ref{sec:3} we classify the four-fermion contributions to the Lagrangian involving one or two top quarks. The explicit calculations of top decay widths are presented in section~\ref{sec:4}, cross sections for single top production are given in section~\ref{sec:5} and for top pair production in section~\ref{sec:6}. We summarise our results in section~\ref{sec:7}.

\section{Four-fermion operator basis}
\label{sec:2}

We follow the notation in Refs.~\cite{Buchmuller:1985jz,Arzt:1994gp} for gauge-invariant operators, introducing flavour indices $i,j,k,l = 1,2,3$. The left-handed weak $\text{SU}(2)_L$ doublets are $q_{Li}$, $\ell_{Li}$ and the right-handed singlets $u_{Ri}$, $d_{Ri}$, $e_{Ri}$. The Pauli matrices are $\tI$, $I=1,2,3$, the Gell-Mann matrices $\la$, $a=1,\dots,8$, normalised to $\text{tr}(\la \lambda^b) = 2 \delta_{ab}$, and $\epsilon=i\tau^2$. Fermion fields are ordered according to their spinorial index contraction. In operators with four quark fields, the subindices $a$, $b$ indicate the pairs with colour indices contracted, if this pairing is different from the one for the spinorial contraction. Our basis consists of the following operators:

\noindent
(i) $\bar L L \bar L L$ operators
\begin{align}
& O_{qq}^{ijkl} = \oh (\bar q_{Li} \gM q_{Lj}) (\bar q_{Lk} \gm q_{Ll}) \,,
&& O_{qq'}^{ijkl} = \oh (\bar q_{Lia} \gM q_{Ljb}) (\bar q_{Lkb} \gm q_{Lla}) \,,
\notag \\
& O_{\ell q}^{ijkl} = (\bar \ell_{Li} \gM \ell_{Lj}) (\bar q_{Lk} \gm q_{Ll}) \,,
&& O_{\ell q'}^{ijkl} = (\bar \ell_{Li} \gM q_{Lj}) (\bar q_{Lk} \gm \ell_{Ll}) \,,
\notag \\
& O_{\ell \ell}^{ijkl} = \oh (\bar \ell_{Li} \gM \ell_{Lj}) (\bar \ell_{Lk} \gm \ell_{Ll}) \,.
\label{ec:LLLL}
\end{align}
(ii) $\bar R R \bar R R$ operators
\begin{align}
& O_{uu}^{ijkl} = \oh (\bar u_{Ri} \gM u_{Rj}) (\bar u_{Rk} \gm u_{Rl}) \,,
&& O_{dd}^{ijkl} = \oh (\bar d_{Ri} \gM d_{Rj}) (\bar d_{Rk} \gm d_{Rl}) \,,
\notag \\
& O_{ud}^{ijkl} = (\bar u_{Ri} \gM u_{Rj}) (\bar d_{Rk} \gm d_{Rl}) \,,
&& O_{ud'}^{ijkl} = (\bar u_{Ria} \gM u_{Rjb}) (\bar d_{Rkb} \gm d_{Rla}) \,,
\notag \\
& O_{eu}^{ijkl} = (\bar e_{Ri} \gM e_{Rj}) (\bar u_{Rk} \gm u_{Rl}) \,, 
&& O_{ed}^{ijkl} = (\bar e_{Ri} \gM e_{Rj}) (\bar d_{Rk} \gm d_{Rl}) \,,
\notag \\
& O_{ee}^{ijkl} = \oh (\bar e_{Ri} \gM e_{Rj}) (\bar e_{Rk} \gm e_{Rl}) \,.
\label{ec:RRRR}
\end{align}
(iii) $\bar L R \bar R L$ operators
\begin{align}
& O_{qu}^{ijkl} = (\bar q_{Li} u_{Rj}) (\bar u_{Rk} q_{Ll}) \,,
&& O_{qu'}^{ijkl} = (\bar q_{Lia} u_{Rjb}) (\bar u_{Rkb} q_{Lla}) \,,
\notag \\
& O_{qd}^{ijkl} = (\bar q_{Li} d_{Rj}) (\bar d_{Rk} q_{Ll}) \,,
&& O_{qd'}^{ijkl} = (\bar q_{Lia} d_{Rjb}) (\bar d_{Rkb} q_{Lla}) \,,
\notag \\
& O_{\ell u}^{ijkl} = (\bar \ell_{Li} u_{Rj}) (\bar u_{Rk} \ell_{Ll}) \,,
&& O_{\ell d}^{ijkl} = (\bar \ell_{Li} d_{Rj}) (\bar d_{Rk} \ell_{Ll}) \,,
\notag \\
& O_{qe}^{ijkl} = (\bar q_{Li} e_{Rj}) (\bar e_{Rk} q_{Ll}) \,,
&& O_{qde}^{ijkl} = (\bar \ell_{Li} e_{Rj}) (\bar d_{Rk} q_{Ll}) \,,
\notag \\
& O_{\ell e}^{ijkl} = (\bar \ell_{Li} e_{Rj}) (\bar e_{Rk} \ell_{Ll}) \,.
\label{ec:LRRL}
\end{align}
(iv) $\bar L R \bar L R$ operators
\begin{align}
& O_{qq\epsilon}^{ijkl} = (\bar q_{Li} u_{Rj}) \left[ (\bar q_{Lk} \epsilon)^T d_{Rl} \right] \,,
&& O_{qq\epsilon'}^{ijkl} = (\bar q_{Lia} u_{Rjb}) \left[ (\bar q_{Lkb} \epsilon)^T d_{Rla} \right] \,,
\notag \\
& O_{\ell q\epsilon}^{ijkl} = (\bar \ell_{Li} e_{Rj}) \left[ (\bar q_{Lk} \epsilon)^T u_{Rl} \right] \,,
&& O_{q \ell \epsilon}^{ijkl} = (\bar q_{Li} e_{Rj}) \left[ (\bar \ell_{Lk} \epsilon)^T u_{Rl} \right] \,.
\label{ec:LRLR}
\end{align}
All the remaining four-fermion operators written in Refs.~\cite{Buchmuller:1985jz,Arzt:1994gp} but not included in our list can be written in terms of these, using the completeness relations for Pauli and Gell-Mann matrices
\begin{align}
& \sum_{I=1}^3 (\tI)_{ij} (\tI)_{kl} = 2 \left( \delta_{il} \delta_{kj} - \oh \delta_{ij} \delta_{kl} \right) \,, \notag \\
& \sum_{a=1}^8 (\la)_{ij} (\la)_{kl} = 2 \left( \delta_{il} \delta_{kj} - \ot \delta_{ij} \delta_{kl} \right) \,,
\label{ec:comp}
\end{align}
and Fierz rearrangements
\begin{eqnarray}
(\bar A_L \gM B_L) (\bar C_L \gm D_L) & = & (\bar A_L \gM D_L) (\bar C_L \gm B_L) \,, \notag \\
(\bar A_R \gM B_R) (\bar C_R \gm D_R) & = & (\bar A_R \gM D_R) (\bar C_R \gm B_R) \,, \notag \\
(\bar A_R \gM B_R) (\bar C_L \gm D_L) & = & -2 (\bar C_L B_R) (\bar A_R D_L) \,,
\end{eqnarray}
where $A,B,C,D$ are four-component spinors of the chirality indicated in each case. Explicitly, the operators written in Refs.~\cite{Buchmuller:1985jz,Arzt:1994gp} but missing from our list are
\begin{align}
& O_{\ell \ell}^{(3,ijkl)} = \oh (\bar \ell_{Li} \gm \tI \ell_{Lj}) (\bar \ell_{Lk} \gM \tI \ell_{Ll})
= 2 O_{\ell \ell}^{ilkj}-O_{\ell \ell}^{ijkl}
\,, \notag \\
& O_{qq}^{(8,1,ijkl)} = \oh (\bar q_{Li} \gm \la q_{Lj}) (\bar q_{Lk} \gM \la q_{Ll})
= 2 O_{qq'}^{ijkl} - \twt O_{qq}^{ijkl}
\,, \notag \\
& O_{qq}^{(1,3,ijkl)} = \oh (\bar q_{Li} \gm \tI q_{Lj}) (\bar q_{Lk} \gM \tI q_{Ll})
= 2 O_{qq'}^{ilkj} - O_{qq}^{ijkl}
\,, \notag \\
& O_{qq}^{(8,3,ijkl)} = \oh (\bar q_{Li} \gm \la \tI q_{Lj}) (\bar q_{Lk} \gM \la \tI q_{Ll})
= 4 O_{qq}^{ilkj} + \twt O_{qq}^{ijkl} - 2 O_{qq'}^{ijkl} - \ft O_{qq'}^{ilkj}
\,, \notag  \\
& O_{\ell q}^{(3,ijkl)} = (\bar \ell_{Li} \gm \tI \ell_{Lj}) (\bar q_{Lk} \gM \tI q_{Ll})
= 2 O_{\ell q'}^{ilkj} - O_{\ell q}^{ijkl}
\,, \notag \\
& O_{uu}^{(8,ijkl)} = \oh (\bar u_{Ri} \gm \la u_{Rj}) (\bar u_{Rk} \gM \la u_{Rl})
= 2 O_{uu}^{ilkj} - \twt O_{uu}^{ijkl}
\,, \notag  \\
& O_{dd}^{(8,ijkl)} = \oh (\bar d_{Ri} \gm \la d_{Rj}) (\bar d_{Rk} \gM \la d_{Rl})
= 2 O_{dd}^{ilkj} - \twt O_{dd}^{ijkl}
\,, \notag \\
& O_{ud}^{(8,ijkl)} = (\bar u_{Ri} \gm \la u_{Rj}) (\bar d_{Rk} \gM \la d_{Rl})
= 2 O_{ud'}^{ilkj} - \twt O_{ud}^{ijkl}
\,, \notag \\
& O_{qu}^{(8,ijkl)} = (\bar q_{Li} \la u_{Rj}) (\bar u_{Rk} \la q_{Ll})
= 2 O_{qu'}^{ijkl} - \twt O_{qu}^{ijkl}
\,, \notag \\
& O_{qd}^{(8,ijkl)} = (\bar q_{Li} \la d_{Rj}) (\bar d_{Rk} \la q_{Ll}) \,,
= 2 O_{qd'}^{ijkl} - \twt O_{qd}^{ijkl}
\,, \notag \\
& O_{qq\epsilon}^{(8,ijkl)} = (\bar q_{Li} \la u_{Rj}) \left[ (\bar q_{Lk} \epsilon)^T \la d_{Rl} \right]
= 2 O_{qq\epsilon'}^{ijkl} - \twt O_{qq\epsilon}^{ijkl} \,.
\end{align}
Some of these relations have previously been obtained in Ref.~\cite{Nomura:2009tw}.
In summary: in our basis we have (i) dropped from the list in Refs.~\cite{Buchmuller:1985jz,Arzt:1994gp} the unnecessary operators $O_{\ell \ell}^{(3,ijkl)}$, $O_{qq}^{(1,3,ijkl)}$, $O_{qq}^{(8,3,ijkl)}$, $O_{uu}^{(8,ijkl)}$ and $O_{dd}^{(8,ijkl)}$; (ii) replaced six operators,
\begin{eqnarray}
O_{qq}^{(8,1,ijkl)} & \to & O_{qq'}^{ijkl} = \ot O_{qq}^{ijkl} + \oh O_{qq}^{(8,1,ijkl)}
\,, \notag \\
O_{\ell q}^{(3,ijkl)} & \to & O_{\ell q'}^{ijkl} = \oh O_{\ell q}^{ilkj} + \oh O_{\ell q}^{(3,ilkj)}
\,, \notag \\
O_{ud}^{(8,ijkl)} & \to & O_{ud'}^{ijkl} = \ot O_{ud}^{ilkj} + \oh O_{ud}^{(8,ilkj)}
\,, \notag \\
O_{qu}^{(8,ijkl)} & \to & O_{qu'}^{ijkl} = \ot O_{qu}^{ijkl} + \oh O_{qu}^{(8,ijkl)}
\,, \notag \\
O_{qd}^{(8,ijkl)} & \to & O_{qd'}^{ijkl} = \ot O_{qd}^{ijkl} + \oh O_{qd}^{(8,ijkl)}
\,, \notag \\
O_{qq\epsilon}^{(8,ijkl)} & \to & O_{qq\epsilon'}^{ijkl} = \ot O_{qq\epsilon}^{ijkl}
+ \oh O_{qq\epsilon}^{(8,ijkl)}
\,.
\end{eqnarray}
We see that these substitutions lead to a larger ``symmetry'' in our basis than in the previous ones, which is apparent with a glance at Eqs.~(\ref{ec:LLLL})--(\ref{ec:LRLR}).
In particular, our operators do not involve $\la$ (nor $\tI$) matrices but instead we have operators $O_{qq'}^{ijkl}$, $O_{ud'}^{ijkl}$, 
$O_{qu'}^{ijkl}$, $O_{qd'}^{ijkl}$ and $O_{qq\epsilon'}^{ijkl}$ in which the colour and spinorial indices are contracted between different quarks pairs. This obviously simplifies amplitude calculations because the $\la_{ij} \la_{kl}$ colour sums do not have to be done case by case. But a more important advantage is that operators with the same quark fields but different colour contractions, {\em e.g.} $O_{ud}^{ijkl}$ and $O_{ud'}^{ijkl}$, correspond to the two possible colour flows in the four-fermion amplitudes and only interfere when all colours are equal. These interferences are trivial (100\% constructive), take the same form in most processes and are easy to parameterise.
The symmetry in our basis leads to simple expressions for top decay widths and production cross sections, as it will be seen in sections~\ref{sec:4}--\ref{sec:6}.

\section{Four-fermion contributions}
\label{sec:3}

In this section we provide a complete list of independent four-fermion operators which give Lagrangian terms with one or two top quarks. We use the shorthand
\begin{equation}
\alpha_x = \frac{C_x}{\Lambda^2}
\end{equation}
to easy the notation, and perform Fierz rearrangements of $\bar L R \bar R L$ terms. We classify the operators according to the four-fermion terms they give, which in turn determine the processes to which they can contribute. We find it also convenient to separate the four-fermion operators giving terms with a $b$ quark (which is the $\text{SU}(2)_L$ partner of the top) from those giving lighter down-type quarks $d_k$, $k=1,2$.

\subsection{Four-fermion terms $t \bar b \bar u_i d_j$, $t \bar t \bar u_i u_j$, $t \bar t \bar d_i d_j$}
\label{sec:3.1}

Four-fermion terms in these three groups arise from four-quark gauge invariant operators with two flavour indices equal to three. Often, the same gauge-invariant operator gives contributions in more than one of these groups. For this reason it is convenient to study them together, allowing for an easy comparison between the different four-fermion contributions. Needless to say, the links between terms in the different groups are due to the gauge symmetry.

The gauge-invariant operators giving four-fermion terms $t \bar b \bar u_i d_j$, $t \bar t \bar u_i u_j$ and $t \bar t \bar d_i d_j$ (plus the Hermitian conjugate), with $u_{i,j} = u,c$, $d_{i,j}=d,s,b$, are collected in Table~\ref{tab:op1}. We also give the number of independent operators in each case. Note that, for example, $O_{qq}^{3123} = O_{qq}^{3213^\dagger}$ and these two operators are not independent. The same applies to other flavour combinations not listed, involving different index ordering.

\begin{table}[htb]
\begin{center}
\begin{tabular}{ccccccccccc}
& $t \bar b \bar u_i d_j$ & $t \bar t \bar u_i u_j$ & $t \bar t \bar d_i d_j$ & \# & \quad \quad && $t \bar b \bar u_i d_j$ & $t \bar t \bar u_i u_j$ & $t \bar t \bar d_i d_j$ & \# \\
\hline
$O_{qq^{(')}}^{3ji3}$   & \ok & \ok & --  & 10 & &
  $O_{qu^{(')}}^{3ji3}$ & --  & \ok & --  & 6 \\
$O_{qq^{(')}}^{ij33}$   & --  & \ok & \ok & 12  & &
  $O_{qd^{(')}}^{ij33}$ & \ok & --  & --  & 12 \\
$O_{uu}^{ij33}$         & --  & \ok & --  & 3 & &
  $O_{qd^{(')}}^{3ji3}$ & --  & --  & \ok & 12 \\
$O_{uu}^{3ji3}$        & --  & \ok & --  & 3 & &
  $O_{qq\epsilon^{(')}}^{i33j}$ & \ok & -- & \ok & 12 \\
$O_{ud^{(')}}^{i33j}$   & \ok & --  & --  & 12 & &
  $O_{qq\epsilon^{(')}}^{33ij}$ & \ok & -- & \ok & 12 \\
$O_{ud^{(')}}^{33ij}$  & --  & --  & \ok & 12 & &
  $O_{qq\epsilon^{(')}}^{3ij3}$ & \ok & -- & --  & 12 \\
$O_{qu^{(')}}^{33ij}$  & \ok & \ok & --  & 12 & &
  $O_{qq\epsilon^{(')}}^{ji33}$ & \ok & -- & --  & 8 \\
$O_{qu^{(')}}^{i33j}$  & --  & \ok & \ok & 12
\end{tabular}
\end{center}
\caption{Gauge-invariant operators giving four-fermion terms $t \bar b \bar u_i d_j$, $t \bar t \bar u_i u_j$ and $t \bar t \bar d_i d_j$ (plus the Hermitian conjugate). The number of independent operators is also indicated.}
\label{tab:op1} 
\end{table}

Operators involving $t \bar b \bar u_i d_j$ fields contribute to the top three-body decay
$t \to b u_i \bar d_j$ and processes related by crossing symmetry and/or charge conjugation, such as single top production in hadron collisions, $u_i b \to d_j t$, $\bar d_j b \to \bar u_i t$ and $u_i \bar d_j \to \bar b t$.
For each set of fields $t$, $\bar b$, $\bar u_i$, $d_j$ there are 16 independent four-fermion terms, in 4 vector and 4 scalar structures, each with two possible colour contractions. Symbolically, we have
\begin{align}
& \bar L L \bar L L \,,\quad \bar L L \bar R R \,,\quad \bar R R \bar L L \,,\quad \bar R R \bar R R \,, \notag \\
& \bar L_a L_b \bar L_b L_a \,,\quad \bar L_a L_b \bar R_b R_a \,,\quad \bar R_a R_b L_b L_a \,,\quad \bar R_a R_b \bar R_b R_a \,, \notag \\
& \bar L R \bar L R \,,\quad \bar R L \bar R L \quad \text{(two orderings)} \,, \notag \\
& \bar L_a R_b \bar L_b R_a \,,\quad \bar R_a L_b \bar R_b L_a \quad \text{(two orderings)} \,.
\label{ec:16t}
\end{align}
All the resulting effective Lagrangian terms are collected in Table~\ref{tab:res-bH}, with their corresponding effective operator coefficients. We only show 
the terms involving $t$ fields, the Hermitian conjugate ones $\bar t b u_i \bar d_j$ have the  complex conjugate coefficients.  
In these tables, the coefficient of each four-fermion term in the Lagrangian can be read by simply intersecting the corresponding row and column. In the case of $\bar L R \bar L R$ and $\bar R L \bar R L$ terms the dots stand for the insertion of the two fields in the upper row, in the order specified (their chirality is determined by the fields in the left column, that is, $\bar b_L t = \bar b_L t_R$, $\bar b_R t = \bar b_R t_L$).
The $\bar L L \bar L L$ coefficients, which are not all independent, are shown separately.
The coefficients of Hermitian operators can be assumed real without loss of generality; they are shown over a gray background.

%
%
\begin{table}[h]
\begin{center}
\begin{small}
\begin{tabular}{l|ccc}
\multicolumn{1}{c|}{\large \ding{192}} && $(\bar u_L \gm t_L)$ & $(\bar c_L \gm t_L)$ \\[1mm]
\hline \\[-5mm]
$(\bar b_L \gM d_L)$ && \gris $\aqq^{3113}$ & $\aqq^{3213*}/2$ \\[1mm]
$(\bar b_L \gM s_L)$ && $\aqq^{3213}/2$ & \gris $\aqq^{3223}$ \\[1mm]
$(\bar b_L \gM b_L)$ && $\aqq^{3313}/2$ & $\aqq^{3323}/2$
\end{tabular}
\quad \quad
\begin{tabular}{l|ccc}
\multicolumn{1}{c|}{\large \ding{193}} && $(\bar u_{Lb} \gm t_{La})$ & $(\bar c_{Lb} \gm t_{La})$ \\[1mm]
\hline \\[-5mm]
$(\bar b_{La} \gM d_{Lb})$ && \gris $\aqqp^{3113}$ & $\aqqp^{3213*}/2$ \\[1mm]
$(\bar b_{La} \gM s_{Lb})$ && $\aqqp^{3213}/2$ & \gris $\aqqp^{3223}$ \\[1mm]
$(\bar b_{La} \gM b_{Lb})$ && $\aqqp^{3313}/2$ & $\aqqp^{3323}/2$
\end{tabular}
\vspace{5mm}

\begin{tabular}{l|cc}
 & $(\bar u_{Li} \gm t_L)$ & $(\bar u_{Ri} \gm t_R)$ \\[1mm]
\hline \\[-5mm]
$(\bar b_L \gM d_{Lj})$ & {\large \ding{192}} & $-\aqup^{33ij}/2$ \\[1mm]
$(\bar b_R \gM d_{Rj})$ & $-\aqdp^{ij33}/2$ & $\aud^{i33j}$
\end{tabular}
\quad \quad
\begin{tabular}{l|cc}
 & $(\bar u_{Lib} \gm t_{La})$ & $(\bar u_{Rib} \gm t_{Ra})$ \\[1mm]
\hline \\[-5mm]
$(\bar b_{La} \gM d_{Ljb})$ & {\large \ding{193}} & $-\aqu^{33ij}/2$ \\[1mm]
$(\bar b_{Ra} \gM d_{Rjb})$ & $-\aqd^{ij33}/2$ & $\audp^{i33j}$
\end{tabular}
\vspace{5mm}

\begin{tabular}{l|cccc}
& $d_j~t$ & $t~d_j$ \\[1mm]
\hline \\[-5mm]
$(\bar b_L \, \cdot \, ) \; (\bar u_{Li} \, \cdot \, )$ 
  & $-\aqqe^{i33j}$ & $\aqqe^{33ij}$ \\[1mm]
$(\bar b_R \, \cdot \, ) \; (\bar u_{Ri} \, \cdot \, )$ 
  & $-\aqqe^{3ij3*}$ & $\aqqe^{ji33*}$
\end{tabular}
\quad \quad
\begin{tabular}{l|cccc}
& $d_{jb}~t_a$ & $t_b~d_{ja}$ \\[1mm]
\hline \\[-5mm]
$(\bar b_{La} \, \cdot \, ) \; (\bar u_{Lib} \, \cdot \, )$ 
  & $-\aqqep^{i33j}$ & $\aqqep^{33ij}$ \\[1mm]
$(\bar b_{Ra} \, \cdot \, ) \; (\bar u_{Rib} \, \cdot \, )$ 
  & $-\aqqep^{3ij3*}$ & $\aqqep^{ji33*}$
\end{tabular}
\end{small}
\end{center}
\caption{Four-fermion contributions with $t \bar b \bar u_i d_j$ fields, being
$i=1,2$, $j=1,2,3$. For $\bar L R \bar L R$ and $\bar R L \bar R L$ terms the dots stand for the insertion of the fields in the upper row. Real coefficients are shown over a gray background.}
\label{tab:res-bH} 
\end{table}

Operators involving $t \bar t \bar u_i u_j$ fields contribute for example to top pair production in hadron collisions, $\bar u_i u_j \to t \bar t$. There are 10 independent 
four-fermion terms, all of vector type, with two possible colour contractions,
\begin{align}
& \bar L L \bar L L \,,\quad \bar R R \bar R R \,, \notag \\
& \bar L L \bar R R \,,\quad \bar R R \bar L L \quad \text{(three terms)} \,, \notag \\
& \bar L_a L_b \bar L_b L_a \,,\quad \bar R_a R_b \bar R_b R_a \,, \notag \\
& \bar L_a L_b \bar R_b R_a \,,\quad \bar R_a R_b \bar L_b L_a \quad \text{(three terms)} \,.
\end{align}
The relevant four-fermion terms are collected in Table~\ref{tab:res-uutt}. Notice that many of them are not independent, but related by Hermitian conjugation.
In the upper and middle tables the dots stand for the insertion of the fields in the second row (the chirality of the latter is determined by the fields in the left column); the resulting bilinear multiplies the corresponding one in the first row.
The coefficients of Hermitian operators can be assumed to be real and are shown over a gray background.

%
%
\begin{table}[htb]
\begin{center}
\begin{small}
\begin{tabular}{l|ccccc}
&& \multicolumn{2}{c}{$\otimes (\bar t_L \gm t_L)$} & \multicolumn{2}{c}{$\otimes (\bar t_R \gm t_R)$} \\[-2mm]
&& $u$ & $c$ & $u$ & $c$ \\[1mm]
\hline \\[-5mm]
$(\bar u_L \gM \, \cdot \, )$
 && \gris $\aqq^{1133} + \aqqp^{3113}$ & $(\aqq^{1233}+\aqqp^{3213})/2$
  & \gris $-\aqup^{1331}$ & $-\aqup^{1332}/2$ \\[1mm]
$(\bar c_L \gM \, \cdot \, )$
 && $(\aqq^{1233*}+\aqqp^{3213*})/2$     & \gris $\aqq^{2233} + \aqqp^{3223}$
  & $-\aqup^{1332*}/2$  & \gris $-\aqup^{2332}$ \\[1mm]
$(\bar u_R \gM \, \cdot \, )$
 &&  \gris $-\aqup^{3113}$ & $-\aqup^{3213}/2$
  & \gris $\auu^{1133}$ & $\auu^{1233}/2$ \\[1mm]
$(\bar c_R \gM \, \cdot \, )$ 
 && $-\aqup^{3213*}/2$ & \gris $-\aqup^{3223}$
  & $\auu^{1233*}/2$ & \gris $\auu^{2233}$ \\[1mm]
\end{tabular}
\vspace{5mm}

\begin{tabular}{l|ccccc}
&& \multicolumn{2}{c}{$\otimes (\bar t_{Lb} \gm t_{La})$} & \multicolumn{2}{c}{$\otimes (\bar t_{Rb} \gm t_{Ra})$} \\[-2mm]
&& $u_b$ & $c_b$ & $u_b$ & $c_b$ \\[1mm]
\hline \\[-5mm]
$(\bar u_{La} \gM \, \cdot \, )$
 && \gris $\aqqp^{1133} + \aqq^{3113}$ & $(\aqqp^{1233}+\aqq^{3213})/2$
  & \gris $-\aqu^{1331}$ & $-\aqu^{1332}/2$ \\[1mm]
$(\bar c_{La} \gM \, \cdot \, )$
 && $(\aqqp^{1233*}+\aqq^{3213*})/2$     & \gris $\aqqp^{2233} + \aqq^{3223}$
  & $-\aqu^{1332*}/2$  & \gris $-\aqu^{2332}$ \\[1mm]
$(\bar u_{Ra} \gM \, \cdot \, )$
 && \gris $-\aqu^{3113}$ & $-\aqu^{3213}/2$
  & \gris $\auu^{3113}$ & $\auu^{3213}/2$ \\[1mm]
$(\bar c_{Ra} \gM \, \cdot \, )$ 
 && $-\aqu^{3213*}/2$ & \gris $-\aqu^{3223}$
  & $\auu^{3213*}/2$ & \gris $\auu^{3223}$ \\[1mm]
\end{tabular}
\vspace{5mm}

\begin{tabular}{l|cc}
 & $(\bar u_{Li} \gm t_L)$ & $(\bar u_{Ri} \gm t_R)$ \\[1mm]
\hline \\[-5mm]
$(\bar t_L \gM u_{Lj})$ & -- & $-\aqup^{33ij}/2$ \\[1mm]
$(\bar t_R \gM u_{Rj})$ & $-\aqup^{33ji*}/2$ & --
\end{tabular}
\quad
\begin{tabular}{l|cc}
 & $(\bar u_{Lib} \gm t_{La})$ & $(\bar u_{Rib} \gm t_{Ra})$ \\[1mm]
\hline \\[-5mm]
$(\bar t_{La} \gM u_{Ljb})$ & -- & $-\aqu^{33ij}/2$ \\[1mm]
$(\bar t_{Ra} \gM u_{Rjb})$ & $-\aqu^{33ji*}/2$ & --
\end{tabular}
\end{small}
\end{center}
\caption{Four-fermion contributions with $t \bar t \bar u_i u_j$ fields, being
$i,j=1,2$. In the upper and middle tables the dots stand for the insertion of the fields in the second row; a multiplication by the corresponding bilinear in the first row is understood. Real coefficients are shown over a gray background.}
\label{tab:res-uutt} 
\end{table}

We point out that we have used Fierz identities to rewrite some terms such as $(\bar t_L \gM u_{Lj}) (\bar u_{Li} \gm t_L)$, which arise from independent gauge-invariant operators, in order to have as few different four-fermion structures as possible. Thus, the number of independent four-fermion terms is 10, while the number of operator coefficients is 12. 

Operators involving $t \bar t \bar d_i d_j$ fields also contribute to top pair production, $\bar d_i d_j \to t \bar t$. There are 16 independent 
four-fermion terms, 8 of vector and 8 of scalar type, with two possible colour contractions, as in Eqs.~(\ref{ec:16t}). The relevant four-fermion terms and their coefficients are collected in Table~\ref{tab:res-ddtt}. As in the previous examples,
the dots stand for the insertion of the field(s) in the upper rows and the multiplication by the corresponding bilinear, if so indicated.
The coefficients of Hermitian operators can be assumed real and are shown over a gray background.

%
%
\begin{table}[htb]
\begin{center}
\begin{small}
\begin{tabular}{l|ccccccc}
&& \multicolumn{3}{c}{$\otimes (\bar t_L \gm t_L)$} & \multicolumn{3}{c}{$\otimes (\bar t_R \gm t_R)$} \\[-2mm]
&& $d$ & $s$ & $b$ & $d$ & $s$ & $b$ \\[1mm]
\hline \\[-5mm]
$(\bar d_L \gM \, \cdot \, )$
 && \gris $\aqq^{1133}$    & $\aqq^{1233}/2$      & $\aqq^{1333}/2$
  & \gris $-\aqup^{1331}$  & $-\aqup^{1332}/2$    & $-\aqup^{1333}/2$ \\[1mm]
$(\bar s_L \gM \, \cdot \, )$
 && $\aqq^{1233*}/2$     & \gris $\aqq^{2233}$    & $\aqq^{2333}/2$
  & $-\aqup^{1332*}/2$   & \gris $-\aqup^{2332}$  & $-\aqup^{2333}/2$ \\[1mm]
$(\bar b_L \gM \, \cdot \, )$
 && $\aqq^{1333*}/2$     & $\aqq^{2333*}/2$     & \gris $\aqq^{3333}$
  & $-\aqup^{1333*}/2$   & $-\aqup^{2333*}/2$   & \gris $-\aqup^{3333}$ \\[1mm]
$(\bar d_R \gM \, \cdot \, )$
 && \gris $-\aqdp^{3113}$  & $-\aqdp^{3213}/2$    & $-\aqdp^{3313}/2$
  & \gris $2\,\aud^{3311}$ & $\aud^{3312}$        & $\aud^{3313}$ \\[1mm]
$(\bar s_R \gM \, \cdot \, )$ 
 && $-\aqdp^{3213*}/2$   & \gris $-\aqdp^{3223}$  & $-\aqdp^{3323}/2$
  & $\aud^{3312*}$       & \gris $2\,\aud^{3322}$ & $\aud^{3323}$ \\[1mm]
$(\bar b_R \gM \, \cdot \, )$
 && $-\aqdp^{3313*}/2$   & $-\aqdp^{3323*}/2$   & \gris $-\aqdp^{3333}$
  & $\aud^{3313*}$       & $\aud^{3323*}$       & \gris $2\,\aud^{3333}$ \\[1mm]
\end{tabular}
\vspace{5mm}

\begin{tabular}{l|ccccccc}
&& \multicolumn{3}{c}{$\otimes (\bar t_{Lb} \gm t_{La})$} & \multicolumn{3}{c}{$\otimes (\bar t_{Rb} \gm t_{Ra})$} \\[-2mm]
&& $d_b$ & $s_b$ & $b_b$ & $d_b$ & $s_b$ & $b_b$ \\[1mm]
\hline \\[-5mm]
$(\bar d_{La} \gM \, \cdot \, )$
 && \gris $\aqqp^{1133}$    & $\aqqp^{1233}/2$      & $\aqqp^{1333}/2$
  & \gris $-\aqu^{1331}$  & $-\aqu^{1332}/2$    & $-\aqu^{1333}/2$ \\[1mm]
$(\bar s_{La} \gM \, \cdot \, )$
 && $\aqqp^{1233*}/2$     & \gris $\aqqp^{2233}$    & $\aqqp^{2333}/2$
  & $-\aqu^{1332*}/2$   & \gris $-\aqu^{2332}$  & $-\aqu^{2333}/2$ \\[1mm]
$(\bar b_{La} \gM \, \cdot \, )$
 && $\aqqp^{1333*}/2$     & $\aqqp^{2333*}/2$     & \gris $\aqqp^{3333}$
  & $-\aqu^{1333*}/2$   & $-\aqu^{2333*}/2$   & \gris $-\aqu^{3333}$ \\[1mm]
$(\bar d_{Ra} \gM \, \cdot \, )$
 && \gris $-\aqd^{3113}$  & $-\aqd^{3213}/2$    & $-\aqd^{3313}/2$
  & \gris $2\,\audp^{3311}$ & $\audp^{3312}$        & $\audp^{3313}$ \\[1mm]
$(\bar s_{Ra} \gM \, \cdot \, )$ 
 && $-\aqd^{3213*}/2$   & \gris $-\aqd^{3223}$  & $-\aqd^{3323}/2$
  & $\audp^{3312*}$       & \gris $2\,\audp^{3322}$ & $\audp^{3323}$ \\[1mm]
$(\bar b_{Ra} \gM \, \cdot \, )$
 && $-\aqd^{3313*}/2$   & $-\aqd^{3323*}/2$   & \gris $-\aqd^{3333}$
  & $\audp^{3313*}$       & $\audp^{3323*}$       & \gris $2\,\audp^{3333}$ \\[1mm]
\end{tabular}
\vspace{5mm}

\begin{tabular}{l|cc}
& $d_j~t$ & $t~d_j$ \\[1mm]
\hline \\[-5mm]
$(\bar d_{Li} \, \cdot \, ) \; (\bar t_{L} \, \cdot \, )$ 
  & $-\aqqe^{33ij}$ & $\aqqe^{i33j}$ \\[1mm]
$(\bar d_{Ri} \, \cdot \, ) \; (\bar t_{R} \, \cdot \, )$ 
  & $-\aqqe^{33ji*}$ & $\aqqe^{j33i*}$
\end{tabular}
\quad \quad
\begin{tabular}{l|cc}
& $d_{jb}~t_a$ & $t_b~d_{ja}$ \\[1mm]
\hline \\[-5mm]
$(\bar d_{Lia} \, \cdot \, ) \; (\bar t_{Lb} \, \cdot \, )$ 
  & $-\aqqep^{33ij}$ & $\aqqep^{i33j}$ \\[1mm]
$(\bar d_{Ria} \, \cdot \, ) \; (\bar t_{Rb} \, \cdot \, )$ 
  & $-\aqqep^{33ji*}$ & $\aqqep^{j33i*}$
\end{tabular}
\end{small}
\end{center}
\caption{Four-fermion contributions with $t \bar t \bar d_i d_j$ fields,
being $i,j=1,2,3$. For vector terms the dots stand for the insertion of the fields in the second row; a multiplication by the corresponding bilinear in the first row is understood.
For $\bar L R \bar L R$ and $\bar R L \bar R L$ terms the dots stand for the insertion of the fields in the upper row. Real coefficients are shown over a gray background.}
\label{tab:res-ddtt}
\end{table}

\subsection{Four-fermion terms $t \bar b e_i \bar \nu_j$, $t \bar t e_i \bar e_j$,
$t \bar t \nu_i \bar \nu_j$}
\label{sec:3.2}

These four-fermion terms arise from gauge invariant operators with two quarks and two leptons, with the two quark flavour indices equal to three and the lepton indices arbitrary. The list of gauge-invariant operators and the type(s) of terms they give is presented in Table~\ref{tab:op2}, including the number of independent operators in each case. 

\begin{table}[htb]
\begin{center}
\begin{tabular}{ccccccccccc}
& $t \bar b e_i \bar \nu_j$ & $t \bar t e_i \bar e_j$ & $t \bar t \nu_i \bar \nu_j$ & \# & \quad \quad && $t \bar b e_i \bar \nu_j$ & $t \bar t e_i \bar e_j$ & $t \bar t \nu_i \bar \nu_j$ & \# \\
\hline
$O_{\ell q}^{ji33}$            & --  & \ok & \ok & 6 & &
  $O_{qe}^{3ij3}$              & --  & \ok & --  & 6 \\
$O_{\ell q'}^{j33i}$           & \ok & --  & \ok & 6 & &
  $O_{qde}^{ji33}$             & \ok & --  & --  & 9 \\
$O_{eu}^{ji33}$                & --  & \ok & --  & 6 & &
  $O_{\ell q \epsilon}^{ji33}$ & \ok & \ok & --  & 9 \\
$O_{\ell u}^{j33i}$            & --  & \ok & \ok & 6 & &  
  $O_{q\ell \epsilon}^{3ij3}$  & \ok & \ok & --  & 9
\end{tabular}
\end{center}
\caption{Gauge-invariant operators giving four-fermion terms $t \bar b e_i \bar \nu_j$, $t \bar t e_i \bar e_j$ and $t \bar t \nu_i \bar \nu_j$ (plus the Hermitian conjugate). The number of independent operators is also indicated.}
\label{tab:op2} 
\end{table}

Four-fermion terms with fields $t \bar b e_i \bar \nu_j$ contribute to three-body top decays $t \to b e_i^+ \nu_j$, being $i,j=1,2,3$. Because $\nu_R$ fields are not introduced, there are only four Lorentz structures, two of vector and two of scalar type,
\begin{align}
& \bar L L \bar L L \,,\quad \bar R R \bar L L \,, \notag \\
& \bar L R \bar L R \quad \text{(two orderings)} \,.
\label{ec:4t}
\end{align}
The contributions to the effective Lagrangian are the ones in Table~\ref{tab:res-bL}, plus the Hermitian conjugate. The coefficients of Hermitian operators can be assumed real without loss of generality. They are displayed over a gray background.

%
%
\begin{table}[htb]
\begin{center}
\begin{small}
\begin{tabular}{l|cccc}
\multicolumn{1}{c|}{\large \ding{194}} && $(\bar \nu_{eL} \gm t_L)$ & $(\bar \nu_{\mu L} \gm t_L)$
 & $(\bar \nu_{\tau L} \gm t_L)$ \\[1mm]
\hline \\[-5mm]
$(\bar b_L \gM e_L)$ && \gris $2\,\alqp^{1331}$ & $\alqp^{2331}$
& $\alqp^{3331}$ \\[1mm]
$(\bar b_L \gM \mu_L)$ && $\alqp^{2331*}$ & \gris $2 \,\alqp^{2332}$ 
& $\alqp^{3332}$ \\[1mm]
$(\bar b_L \gM \tau_L)$ && $\alqp^{3332*}$  & $\alqp^{3332*}$
& \gris $2 \,\alqp^{3333}$ \\[1mm]
\end{tabular}
\quad \quad
\begin{tabular}{c}
\begin{tabular}{l|c}
 & $(\bar \nu_{Lj} \gm t_L)$ \\[1mm]
\hline \\[-5mm]
$(\bar b_L \gM e_{Li})$ & {\large \ding{194}} \\[1mm]
$(\bar b_R \gM e_{Ri})$ & $-\aqde^{ji33}/2$
\end{tabular}
\\[12mm]

\begin{tabular}{l|cccc}
& $e_i~t$ & $t~e_i$ \\[1mm]
\hline \\[-5mm]
$(\bar b_L \, \cdot \, ) \; (\bar \nu_{Lj} \, \cdot \, )$ 
  & $\aqle^{3ij3}$ & $-\alqe^{ji33}$
\end{tabular}
\end{tabular}
\end{small}
\end{center}
\caption{Four-fermion contributions with $t \bar b e_i \bar \nu_j$ fields,
with $i,j=1,2,3$. For $\bar L R \bar L R$ terms the dots stand for the insertion of the fields in the upper row. Real coefficients are shown over a gray background.}
\label{tab:res-bL} 
\end{table}

Lagrangian terms with fields $t \bar t e_i \bar e_j$ are involved for example in top pair production at a future linear collider, $e^+ e^- \to t \bar t$. These terms arise in eight possible Lorentz structures, four vector and four scalar terms,
\begin{align}
& \bar L L \bar L L \,,\quad \bar L L \bar R R \,,\quad \bar R R \bar L L \,,\quad \bar R R \bar R R \,, \notag \\
& \bar L R \bar L R \,,\quad \bar R L \bar R L \quad \text{(two orderings)} \,.
\label{ec:8t}
\end{align}
All contributions to the effective Lagrangian are collected in Table~\ref{tab:res-eett}. The coefficients of Hermitian operators are shown over a gray background.

\begin{table}[htb]
\begin{center}
\begin{small}
\begin{tabular}{l|ccccccc}
&& \multicolumn{3}{c}{$\otimes (\bar t_L \gm t_L)$} & \multicolumn{3}{c}{$\otimes (\bar t_R \gm t_R)$} \\[-2mm]
&& $e$ & $\mu$ & $\tau$ & $e$ & $\mu$ & $\tau$ \\[1mm]
\hline \\[-5mm]
$(\bar e_L \gM \,\cdot\,)$
 && \gris $2\,\alq^{1133}$ & $\alq^{2133*}$       & $\alq^{3133*}$
  & \gris $-\alu^{1331}$   & $-\alu^{2331*}/2$    & $-\alu^{3331*}/2$ \\[1mm]
$(\bar \mu_L \gM \,\cdot\,)$
 && $\alq^{2133}$        & \gris $2\,\alq^{2233}$ & $\alq^{3233*}$
  & $-\alu^{2331}/2$    & \gris $-\alu^{2332}$   & $-\alu^{3332*}/2$ \\[1mm]
$(\bar \tau_L \gM \,\cdot\,)$
 && $\alq^{3133}$        & $\alq^{3233}$        & \gris $2\,\alq^{3333}$
  & $-\alu^{3331}/2$     & $-\alu^{3332}/2$     & \gris $-\alu^{3333}$ \\[1mm]
$(\bar e_R \gM \,\cdot\,)$
 && \gris $-\aqe^{3113}$   & $-\aqe^{3123*}/2$    & $-\aqe^{3133*}/2$
  & \gris $2\,\aeu^{1133}$ & $\aeu^{2133*}$       & $\aeu^{3133*}$ \\[1mm]
$(\bar \mu_R \gM \,\cdot\,)$
 && $-\aqe^{3123}/2$     & \gris $-\aqe^{3223}$   & $-\aqe^{3233*}/2$
  & $\aeu^{2133}$        & \gris $2\,\aeu^{2233}$ & $\aeu^{3233*}$ \\[1mm]
$(\bar \tau_R \gM \,\cdot\,)$
 && $-\aqe^{3133}/2$     & $-\aqe^{3233}/2$     & \gris $-\aqe^{3333}$
  & $\aeu^{3133}$        & $\aeu^{3233}$        & \gris $2\,\aeu^{3333}$ \\[1mm]
\end{tabular}
\vspace{5mm}

\begin{tabular}{l|cccc}
& $e_i~t$ & $t~e_i$ \\[1mm]
\hline \\[-5mm]
$(\bar t_L \, \cdot \, ) \; (\bar e_{Lj} \, \cdot \, )$ 
  & $-\aqle^{3ij3}$ & $\alqe^{ji33}$ \\[1mm]
$(\bar t_R \, \cdot \, ) \; (\bar e_{Rj} \, \cdot \, )$ 
  & $-\aqle^{3ji3*}$ & $\alqe^{ij33*}$
\end{tabular}
\end{small}
\end{center}
\caption{Four-fermion contributions with $t \bar t e_i \bar e_j$ fields, with $i,j=1,2,3$. For vector terms the dots stand for the insertion of the fields in the second row; a multiplication by the corresponding bilinear in the first row is understood.
For $\bar L R \bar L R$ and $\bar R L \bar R L$ terms the dots stand for the insertion of the fields in the upper row. Real coefficients are shown over a gray background.}
\label{tab:res-eett}
\end{table}

We also give for completeness the $t \bar t \nu_i \bar \nu_j$ terms, although they seem to have little relevance for phenomenology. After using Fierz rearrangements on some terms, there are only two independent structures,
\begin{align}
& \bar L L \bar L L \,,\quad \bar L L \bar R R \,,
\label{ec:2t}
\end{align}
with three independent operator coefficients for each set of fields $t \bar t \nu_i \bar \nu_j$. We give in Table~\ref{tab:res-nunutt} the full set of Lagrangian terms with their corresponding operator coefficients. It is worth pointint out that, despite the fact that these four-fermion terms do not contribute to lowest order processes in hadron or lepton collisions, the operators involved can be probed either in top decays or in top pair production through the $t \bar b e_i \bar \nu_j$ or $t \bar t e_i \bar e_j$ terms generated, see Table~\ref{tab:op2}.

%
%
\begin{table}[htb]
\begin{center}
\begin{small}
\begin{tabular}{l|cccc}
&& \multicolumn{3}{c}{$\otimes (\bar t_L \gm t_L)$} \\[-2mm]
&& $\nu_e$ & $\nu_\mu$ & $\nu_\tau$ \\[1mm]
\hline \\[-5mm]
$(\bar \nu_{eL} \gM \,\cdot\,)$
 && \gris $2 (\alq^{1133} + \alqp^{1331})$
  & $\alq^{2133*}+\alqp^{2331*}$
  & $\alq^{3133*}+\alqp^{3331*}$ \\[0.5mm]
$(\bar \nu_{\mu L} \gM \,\cdot\,)$
 && $\alq^{2133}+\alqp^{2331}$
  & \gris $2(\alq^{2233}+\alqp^{2332})$
  & $\alq^{3233*}+\alqp^{3332*}$ \\[0.5mm]
$(\bar \nu_{\tau L} \gM \,\cdot\,)$ 
 && $\alq^{3133}+\alqp^{3331}$
  & $\alq^{3233}+\alqp^{3332}$
  & \gris $2(\alq^{3333}+\alqp^{3333})$ \\[0.5mm]
\end{tabular}
\vspace{5mm}

\begin{tabular}{l|cccc}
&& \multicolumn{3}{c}{$\otimes (\bar t_R \gm t_R)$} \\[-2mm]
&& $\nu_e$ & $\nu_\mu$ & $\nu_\tau$ \\[1mm]
\hline \\[-5mm]
$(\bar \nu_{eL} \gM \,\cdot\,)$
 && \gris $-\alu^{1331}$   & $-\alu^{2331*}/2$    & $-\alu^{3331*}/2$ \\[0.5mm]
$(\bar \nu_{\mu L} \gM \,\cdot\,)$
 && $-\alu^{2331}/2$    & \gris $-\alu^{2332}$   & $-\alu^{3332*}/2$ \\[0.5mm]
$(\bar \nu_{\tau L} \gM \,\cdot\,)$ 
 && $-\alu^{3331}/2$     & $-\alu^{3332}/2$     & \gris $-\alu^{3333}$ \\[0.5mm]
\end{tabular}
\end{small}
\end{center}
\caption{Four-fermion contributions with $t \bar t \nu_i \bar \nu_j$ fields, with $i,j=1,2,3$. The dots stand for the insertion of the fields in the second row; a multiplication by the corresponding bilinear in the first row is understood. Real coefficients are shown over a gray background.}
\label{tab:res-nunutt}
\end{table}

\subsection{Four-fermion terms $t \bar d_k \bar u_i d_j$, $t \bar u_k \bar u_i u_j$, $t t \bar u_k \bar u_i$}
\label{sec:3.3}

These are four-fermion terms with $d_k = d,s$ (the case $d_k=b$ was presented in section~\ref{sec:3.1}), $u_{i,j} = u,c$, $d_j=d,s,b$. They appear from four-quark gauge invariant operators with one or two flavour indices equal to three. (In the latter case there is no overlap with the ones studied in section~\ref{sec:3.1}.) The gauge-invariant operators giving such terms are collected in Table~\ref{tab:op3}, also including the number of independent operators. Four-fermion terms with two like-sign top quarks appear from the same operators giving $t \bar u_k \bar u_i u_j$ terms, but when $j=3$.
Note that for $O_{qu^{(')}}^{k3ij}$, $O_{qu^{(')}}^{ijk3}$ and $O_{qq\epsilon^{(')}}^{ji3k}$, $O_{qq\epsilon^{(')}}^{3ijk}$ there is some double counting of flavour combinations when $j=3$, and the total number of operators in each case is 40.

\begin{table}[htb]
\begin{center}
\begin{tabular}{ccccccccccc}
& $t \bar d_k \bar u_i d_j$ & $t \bar u_k \bar u_i u_j$ & $t t \bar u_k \bar u_i$
& \# & \quad \quad && $t \bar d_k \bar u_i d_j$ & $t \bar u_k \bar u_i u_j$ & $t t \bar u_k \bar u_i$ & \# \\
\hline
$O_{qq^{(')}}^{kji3}$           & \ok & \ok  & \ok & 22 & &
  $O_{qd^{(')}}^{ijk3}$         & \ok & --   & --  & 24 \\
$O_{uu}^{kji3}$                 & --  & \ok  & \ok & 11  & &
  $O_{qq\epsilon^{(')}}^{i3kj}$ & \ok & --   & --  & 24 \\
$O_{ud^{(')}}^{i3kj}$           & \ok & --   & --  & 24 & &
  $O_{qq\epsilon^{(')}}^{3ijk}$ & \ok & --   & --  & 24 \\
$O_{qu^{(')}}^{k3ij}$           & \ok & \ok  & \ok & 24 & &
  $O_{qq\epsilon^{(')}}^{ji3k}$ & \ok & --   & --  & 16 \\
$O_{qu^{(')}}^{ijk3}$           & --  & \ok  & \ok & 16 & &
\end{tabular}
\end{center}
\caption{Gauge-invariant operators giving four-fermion terms $t \bar d_k \bar u_i d_j$, $t \bar u_k \bar u_i u_j$ and $tt\bar u_k \bar u_i$ (plus the Hermitian conjugate). The number of independent operators is also indicated.}
\label{tab:op3}
\end{table}

Operators involving $t \bar d_k \bar u_i d_j$ fields mediate top three-body decays
$t \to d_k u_i \bar d_j$ and single top production $u_i d_k \to d_j t$, $\bar d_j d_k \to \bar u_i t$ and $u_i \bar d_j \to \bar d_k t$. These processes already take place in the SM but their amplitudes are very suppressed by small CKM mixings $V_{3k}$.
As in the case $d_k = b$, for each set of fields $t$, $\bar d_k$, $\bar u_i$, $d_j$ there are 16 independent four-fermion terms, 8 of vector and 8 of scalar type, with two possible colour contractions, summarised in Eqs.~(\ref{ec:16t}).
The corresponding Lagrangian terms which involve $t$ fields are collected in Table~\ref{tab:res-dH}. The $\bar L L \bar L L$ operator coefficients are shown separately because they are not all independent. In particular, we note that $O_{qq}^{2313}$ and $O_{qq'}^{2313}$ give two different four-fermion terms in each table.

%
%
\begin{table}[h]
\begin{center}
\begin{small}
\begin{tabular}{l|ccc}
\multicolumn{1}{c|}{\large \ding{195}} && $(\bar u_L \gm t_L)$ & $(\bar c_L \gm t_L)$ \\[1mm]
\hline \\[-5mm]
$(\bar d_L \gM d_L)$ && $\aqq^{1113}/2$ & $\aqq^{1123}/2$ \\[1mm]
$(\bar d_L \gM s_L)$ && $\aqq^{1213}/2$ & $\aqq^{1223}/2$ \\[1mm]
$(\bar d_L \gM b_L)$ && $\aqq^{1313}$ & $\aqq^{2313}/2$ \\[1mm]
$(\bar s_L \gM d_L)$ && $\aqq^{2113}/2$ & $\aqq^{2123}/2$ \\[1mm]
$(\bar s_L \gM s_L)$ && $\aqq^{2213}/2$ & $\aqq^{2223}/2$ \\[1mm]
$(\bar s_L \gM b_L)$ && $\aqq^{2313}/2$ & $\aqq^{2323}$ \\[1mm]
\end{tabular}
\quad \quad
\begin{tabular}{l|ccc}
\multicolumn{1}{c|}{\large \ding{196}} && $(\bar u_{Lb} \gm t_{La})$ & $(\bar c_{Lb} \gm t_{La})$ \\[1mm]
\hline \\[-5mm]
$(\bar d_{La} \gM d_{Lb})$ && $\aqqp^{1113}/2$ & $\aqqp^{1123}/2$ \\[1mm]
$(\bar d_{La} \gM s_{Lb})$ && $\aqqp^{1213}/2$ & $\aqqp^{1223}/2$ \\[1mm]
$(\bar d_{La} \gM b_{Lb})$ && $\aqqp^{1313}$ & $\aqqp^{2313}/2$ \\[1mm]
$(\bar s_{La} \gM d_{Lb})$ && $\aqqp^{2113}/2$ & $\aqqp^{2123}/2$ \\[1mm]
$(\bar s_{La} \gM s_{Lb})$ && $\aqqp^{2213}/2$ & $\aqqp^{2223}/2$ \\[1mm]
$(\bar s_{La} \gM b_{Lb})$ && $\aqqp^{2313}/2$ & $\aqqp^{2323}$ \\[1mm]
\end{tabular}

\vspace{5mm}

\begin{tabular}{l|cc}
 & $(\bar u_{Li} \gm t_L)$ & $(\bar u_{Ri} \gm t_R)$ \\[1mm]
\hline \\[-5mm]
$(\bar d_{Lk} \gM d_{Lj})$ & {\large \ding{195}} & $-\aqup^{k3ij}/2$ \\[1mm]
$(\bar d_{Rk} \gM d_{Rj})$ & $-\aqdp^{ijk3}/2$ & $\aud^{i3kj}$
\end{tabular}
\quad \quad
\begin{tabular}{l|cc}
 & $(\bar u_{Lib} \gm t_{La})$ & $(\bar u_{Rib} \gm t_{Ra})$ \\[1mm]
\hline \\[-5mm]
$(\bar d_{Lka} \gM d_{Ljb})$ & {\large \ding{196}} & $-\aqu^{k3ij}/2$ \\[1mm]
$(\bar d_{Rka} \gM d_{Rjb})$ & $-\aqd^{ijk3}/2$ & $\audp^{i3kj}$
\end{tabular}
\vspace{5mm}

\begin{tabular}{l|cc}
& $d_j~t$ & $t~d_j$ \\[1mm]
\hline \\[-5mm]
$(\bar d_{Lk} \, \cdot \, ) \; (\bar u_{Li} \, \cdot \, )$ 
  & $-\aqqe^{i3kj}$ & $\aqqe^{k3ij}$ \\[1mm]
$(\bar d_{Rk} \, \cdot \, ) \; (\bar u_{Ri} \, \cdot \, )$ 
  & $-\aqqe^{3ijk*}$ & $\aqqe^{ji3k*}$
\end{tabular}
\quad \quad
\begin{tabular}{l|cc}
& $d_{jb}~t_a$ & $t_b~d_{ja}$ \\[1mm]
\hline \\[-5mm]
$(\bar d_{Lka} \, \cdot \, ) \; (\bar u_{Lib} \, \cdot \, )$ 
  & $-\aqqep^{i3kj}$ & $\aqqep^{k3ij}$ \\[1mm]
$(\bar d_{Rka} \, \cdot \, ) \; (\bar u_{Rib} \, \cdot \, )$ 
  & $-\aqqep^{3ijk*}$ & $\aqqep^{ji3k*}$
\end{tabular}
\end{small}
\end{center}
\caption{Four-fermion contributions with $t \bar d_k \bar u_i d_j$ fields, being
$i,k=1,2$, $j=1,2,3$. For $\bar L R \bar L R$ and $\bar R L \bar R L$ terms the dots stand for the insertion of the fields in the upper row.}
\label{tab:res-dH}
\end{table}

On the other hand, four-fermion operators giving $t \bar u_k \bar u_i u_j$ terms mediate top FCN decays $t \to u_k u_i \bar u_j$, with $i,j,k=1,2$, as well as several single top production processes such as $u_i u_k \to u_j t$,
$u_i \bar u_j \to \bar u_k t$. Since there are two identical (up to flavour indices)  fields $u_i$, $u_k$, there are only 6 independent four-fermion structures,
\begin{align}
& \bar L L \bar L L \,,\quad \bar L L \bar R R \,,\quad \bar R R \bar L L \,,\quad \bar R R \bar R R \,, \notag \\
& \bar L_a L_b \bar R_b R_a \,,\quad \bar R_a R_b \bar L_b L_a \,.
\end{align}
The resulting Lagrangian terms are presented in Table~\ref{tab:res-uH}, where the hermitian conjugate ones are also understood. We have rewritten the $(\bar u_{Lka} \gM u_{Ljb}) (\bar u_{Lib} \gm t_{La})$ contributions from $O_{qq'}^{kji3}$ using a Fierz rearrangement and included them in the left table.
We point out that there is no analog to $O_{uu}^{kji3}$ with a different colour index contraction: these operators are redundant as we indicated in section~\ref{sec:2}.
It is also worthwhile remarking here that, since there are two light $\bar u$-type fields, in general each term will give two contributions to the amplitudes. This multiplicity will be carefully dealt with in the calculations performed in the following sections.

%
%
\begin{table}[t]
\begin{center}
\begin{small}
\begin{tabular}{l|cc}
 & $(\bar u_{Li} \gm t_L)$ & $(\bar u_{Ri} \gm t_R)$ \\[1mm]
\hline \\[-5mm]
$(\bar u_{Lk} \gM u_{Lj})$ & ($\aqq^{kji3} + \aqqp^{ijk3})/2$ & $-\aqup^{k3ij}/2$ \\[1mm]
$(\bar u_{Rk} \gM u_{Rj})$ & $-\aqup^{ijk3}/2$ & $\auu^{kji3}/2$
\end{tabular}
\quad
\begin{tabular}{l|cc}
 & $(\bar u_{Lib} \gm t_{La})$ & $(\bar u_{Rib} \gm t_{Ra})$ \\[1mm]
\hline \\[-5mm]
$(\bar u_{Lka} \gM u_{Ljb})$ & -- & $-\aqu^{k3ij}/2$ \\[1mm]
$(\bar u_{Rka} \gM u_{Rjb})$ & $-\aqu^{ijk3}/2$ & --
\end{tabular}
\end{small}
\end{center}
\caption{Four-fermion contributions with $t \bar u_k \bar u_i u_j$ fields, with
$i,j,k=1,2$.}
\label{tab:res-uH}
\end{table}

For operators with two like-sign top quarks, charge conservation requires that the two other fields are light up-type quarks $u_i$, $u_k$.
With two identical $t$ fields, for $\bar L L \bar L L$ and $\bar R R \bar R R$ operators the index combinations with $i$ and $k$ interchanged actually correspond to the same operator. There are only four independent structures for them,
\begin{align}
\bar L L \bar L L \,,\quad \bar R R \bar R R \,,\quad \bar L L \bar R R \,,\quad
\bar L_a L_b \bar R_b R_a \,,
\end{align}
as the other possibilities are equivalent due to the presence of two $t$ fields. The resulting terms are collected in Table~\ref{tab:res-like}.
%
%
\begin{table}[b]
\begin{center}
\begin{small}
\begin{tabular}{l|ccccc}
&& $(\bar u_L \gm t_L)$ & $(\bar c_L \gm t_L)$ & $(\bar u_R \gm t_R)$ & $(\bar c_R \gm t_R)$ \\[1mm]
\hline \\[-5mm]
$(\bar u_L \gM t_L )$
 && $(\aqq^{1313}+\aqqp^{1313})/2$ & $(\aqq^{1323}+\aqqp^{1323})/2$
  & $-\aqup^{1313}/2$ & $-\aqup^{1323}/2$ \\[1mm]
$(\bar c_L \gM t_L )$
 && -- & $(\aqq^{2323}+\aqqp^{2323})/2$
  & $-\aqup^{2313}/2$ & $-\aqup^{2323}/2$ \\[1mm]
$(\bar u_R \gM t_R )$
 && -- & -- & $\auu^{1313}/2$ & $\auu^{1323}/2$ \\[1mm]
$(\bar c_R \gM t_R )$
 && -- & -- & -- & $\auu^{2323}/2$
\\[1mm]
\end{tabular}
\vspace{5mm}

\begin{tabular}{l|ccccc}
&& $(\bar u_{Lb} \gm t_{La})$ & $(\bar c_{Lb} \gm t_{La})$ & $(\bar u_{Rb} \gm t_{Ra})$ & $(\bar c_{Rb} \gm t_{Ra})$ \\[1mm]
\hline \\[-5mm]
$(\bar u_{La} \gM t_{Lb} )$
 && -- & -- & $-\aqu^{1313}/2$ & $-\aqu^{1323}/2$ \\[1mm]
$(\bar c_{La} \gM t_{Lb} )$
 && -- & --  & $-\aqu^{2313}/2$ & $-\aqu^{2323}/2$ \\[1mm]
$(\bar u_{Ra} \gM t_{Rb} )$
 && -- & -- & -- & -- \\[1mm]
$(\bar c_{Ra} \gM t_{Rb} )$
 && -- & -- & -- & --
\\[1mm]
\end{tabular}
\end{small}
\end{center}
\caption{Four-fermion contributions with $t t \bar u_i \bar u_k$ fields, with
$i,k=1,2$.}
\label{tab:res-like} 
\end{table}
 In the upper table, the half below the diagonal is identically equal to the one above, with exchange of the two bilinears. In the second table, $\bar L L \bar L L$ terms have been Fierz-rewritten into the first table and $\bar R R \bar R R$ terms are identical to the ones already included there.

\subsection{Four-fermion terms $t \bar d_k e_i \bar \nu_j$, $t \bar u_k e_i \bar e_j$,
$t \bar u_k \nu_i \bar \nu_j$}
\label{sec:3.4}

These four-fermion terms with $k=1,2$ are analogous to the ones with $k = 3$ previously classified in section~\ref{sec:3.2}. They arise from gauge invariant operators with two quarks and two leptons, with only one quark flavour index equal to three and the lepton indices arbitrary. The list of gauge-invariant operators producing these terms can be found in Table~\ref{tab:op4}, including the number of independent operators.

\begin{table}[htb]
\begin{center}
\begin{tabular}{ccccccccccc}
& $t \bar d_k e_i \bar \nu_j$ & $t \bar u_k e_i \bar e_j$ & $t \bar u_k \nu_i \bar \nu_j$ & \# & \quad \quad && $t \bar d_k e_i \bar \nu_j$ & $t \bar u_k e_i \bar e_j$ & $t \bar u_k \nu_i \bar \nu_j$ & \# \\
\hline
$O_{\ell q}^{jik3}$            & --  & \ok & \ok & 18 & &
  $O_{qde}^{jik3}$             & \ok & --  & --  & 18 \\
$O_{\ell q'}^{j3ki}$           & \ok & --  & \ok & 18 & &
  $O_{\ell q \epsilon}^{jik3}$ & \ok & \ok & --  & 18 \\
$O_{eu}^{jik3}$                & --  & \ok & --  & 18 & &
  $O_{\ell q \epsilon}^{ij3k}$ & --  & \ok & --  & 18 \\
$O_{\ell u}^{j3ki}$            & --  & \ok & \ok & 18 & &  
  $O_{q\ell \epsilon}^{kij3}$  & \ok & \ok & --  & 18 \\
$O_{qe}^{kij3}$                & --  & \ok & --  & 18 & &
$O_{q\ell \epsilon}^{3jik}$    & --  & \ok & --  & 18
\end{tabular}
\end{center}
\caption{Gauge-invariant operators giving four-fermion terms $t \bar d_k e_i \bar \nu_j$, $t \bar u_k e_i \bar e_j$ and $t \bar u_k \nu_i \bar \nu_j$ (plus the Hermitian conjugate). The number of independent operators is also indicated.}
\label{tab:op4} 
\end{table}

Operators with fermion fields $t \bar d_k e_i \bar \nu_j$ can mediate top semileptonic decays $t \to d_k e_i^+ \nu_j$ ($k=1,2$, $i,j=1,2,3$). These decays take place in the SM  when $i=j$, {\em i.e.} if lepton flavour is conserved, but are suppressed by small CKM matrix elements. There are only four possible Lorentz structures, two of vector and two of scalar type, as in Eqs.~(\ref{ec:4t}). The resulting four-fermion contributions are given in Table~\ref{tab:res-dL}, being the Hermitian conjugate terms also present in the Lagrangian.

%
%
\begin{table}[htb]
\begin{center}
\begin{tabular}{l|c}
 & $(\bar \nu_{Lj} \gm t_L)$ \\[1mm]
\hline \\[-5mm]
$(\bar d_{Lk} \gM e_{Li})$ & $\alqp^{j3ki}$ \\[1mm]
$(\bar d_{Rk} \gM e_{Ri})$ & $-\aqde^{jik3}/2$
\end{tabular}
\quad \quad
\begin{tabular}{l|cccc}
& $e_i~t$ & $t~e_i$ \\[1mm]
\hline \\[-5mm]
$(\bar d_{Lk} \, \cdot \, ) \; (\bar \nu_{Lj} \, \cdot \, )$ 
  & $\aqle^{kij3}$ & $-\alqe^{jik3}$
\end{tabular}
\end{center}
\caption{Four-fermion contributions with $t \bar d_k e_i \bar \nu_j$ fields, being $k=1,2$, $i,j=1,2,3$. For $\bar L R \bar L R$ terms the dots stand for the insertion of the fields in the upper row.}
\label{tab:res-dL} 
\end{table}

Four-fermion operators with fields $t \bar u_k e_i \bar e_j$ can mediate top FCN decays 
$t \to u_k e_i^+ e_j^-$ and single top production $e^+ e^- \to t \bar u_k$, absent in the SM at the tree level. There are eight possible Lorentz structures for these four-fermion terms, four of  vector type and four scalar, as in Eqs.~(\ref{ec:8t}).
 All the possible four-fermion terms with their corresponding coefficients are collected in Table~\ref{tab:res-uL}. As before, the Hermitian conjugate terms are also present in the Lagrangian and their coefficients are the complex conjugate of the ones shown.

%
%
\begin{table}[htb]
\begin{center}
\begin{tabular}{l|cc}
 & $(\bar u_{Lk} \gm t_L)$ & $(\bar u_{Rk} \gm t_R)$ \\[1mm]
\hline \\[-5mm]
$(\bar e_{Lj} \gM e_{Li})$ & $\alq^{jik3}$ & $-\alu^{j3ki}/2$ \\[1mm]
$(\bar e_{Rj} \gM e_{Ri})$ & $-\aqe^{kij3}/2$ & $\aeu^{jik3}$
\end{tabular}
\quad \quad
\begin{tabular}{l|cccc}
& $e_i~t$ & $t~e_i$ \\[1mm]
\hline \\[-5mm]
$(\bar u_{Lk} \, \cdot \, ) \; (\bar e_{Lj} \, \cdot \, )$ 
  & $-\aqle^{kij3}$ & $\alqe^{jik3}$ \\[1mm]
$(\bar u_{Rk} \, \cdot \, ) \; (\bar e_{Rj} \, \cdot \, )$ 
  & $-\aqle^{3jik*}$ & $\alqe^{ij3k*}$
\end{tabular}
\end{center}
\caption{Four-fermion contributions with $t \bar u_k e_i \bar e_j$ fields, being $k=1,2$, $i,j=1,2,3$. For $\bar L R \bar L R$ and $\bar R L \bar R L$ terms the dots stand for the insertion of the fields in the upper row.}
\label{tab:res-uL} 
\end{table}

Finally, the $t \bar u_k \nu_i \bar \nu_j$ terms can also mediate top FCN decays
$t \to u_k \nu_i \bar \nu_j$. After using a Fierz rearrangement, there are only two possible Lorentz structures of vector type, as it happens for $t \bar t \nu_i \bar \nu_j$ terms in Eq.~(\ref{ec:2t}). The resulting Lagrangian contributions are collected in Table~\ref{tab:res-tunn}. (The Hermitian conjugate are also understood.)

%
%
\begin{table}[htb]
\begin{center}
\begin{tabular}{l|cc}
 & $(\bar u_{Lk} \gm t_L)$ & $(\bar u_{Rk} \gm t_R)$ \\[1mm]
\hline \\[-5mm]
$(\bar \nu_{Lj} \gM \nu_{Li})$ & $\alq^{jik3}+\alqp^{j3ki}$ & $-\alu^{j3ki}/2$
\end{tabular}
\end{center}
\caption{Four-fermion contributions with $t \bar u_k \nu_i \bar \nu_j$ fields, being $k=1,2$, $i,j=1,2,3$.}
\label{tab:res-tunn} 
\end{table}

\section{Top decay widths}
\label{sec:4}

In this section we calculate and present in turn the partial widths for the several three-body decays mediated by four-fermion operators: charged current decays $t \to d_k u_i \bar d_j$ and $t \to d_k e_i^+ \nu_j$ (where now $d_k=d,s,b$ can be discussed together), and FCN decays $t \to u_k u_i \bar u_j$, $t \to u_k e_i^+ e_j^-$, $t \to u_k \bar \nu_i \nu_j$. For top antiquark decays, the partial widths are the same as the ones shown, but replacing CKM mixing elements and effective operator coefficients by the complex conjugate.
We obtain the top partial widths by integrating the corresponding squared amplitudes over three-body phase space, taking all final state particles massless. For charged current processes a SM contribution, mediated precisely by an on-shell $W$ boson, is present. In these cases we perform the exact integrals including the $W$ boson propagator and then make an expansion in $\Gamma_W/M_W$, keeping the necessary terms. Trace manipulations are done using {\tt FORM}~\cite{Vermaseren:2000nd}.

We only consider interferences between four-fermion operators which do not require chirality flips of light quarks. Also, most interferences between the SM and four-fermion amplitudes are suppressed by a light $u,d,c,s$ quark or lepton mass. The only exception is for decays $t \to d_k u_i \bar b$, where there are some interferences suppressed by $m_b$ which we also neglect. These decays have SM amplitudes already suppressed by small CKM mixings $V_{ub}$ or $V_{cb}$, anyway.

\subsection{$t \to d_k u_i \bar d_j$}
\label{sec:4.1}

This decay can already take place in the SM, with an intermediate on-shell $W$ boson. Integrating in three-body phase space and taking the leading terms in $\Gamma_W/M_W$, we obtain
\begin{eqnarray}
\Gamma_\text{SM} & = & \frac{g^2 m_t}{192 \pi} \left[ \frac{m_t}{M_W} \right]^2 |V_{3k}|^2 |V_{ij}|^2
(1 - 3 x_W^4 + 2 x_W^6) \,,
\end{eqnarray}
with $x_W = M_W/m_t$. This result corresponds to the SM width for $t \to d_k W$ times the branching ratio for $W \to u_i \bar d_j$.

\begin{figure}[htb]
\begin{center}
\epsfig{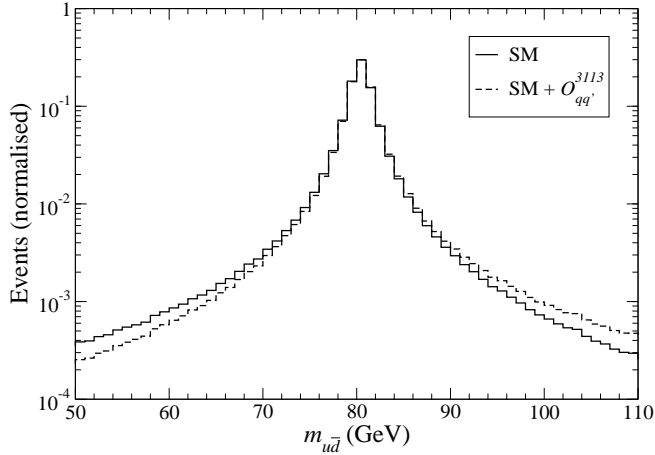}
\caption{Normalised $u \bar d$ invariant mass distribution for the decay $t \to b u \bar d$ within the SM and with $C_{qq'}^{3113} = 10$, $\Lambda = 1$ TeV.}
\label{fig:mw}
\end{center}
\end{figure}

For each set of indices $i,j,k$ there are 16 four-fermion terms in the amplitude, 8 corresponding to the colour flow $t_a \to d_{ka} u_{ib} \bar d_{jb}$ (with $a,b$ colour indices) and 8 for $t_a \to d_{kb} u_{ia} \bar d_{jb}$.
Both sets have a 100\% constructive interference for $a=b$, which happens for one third of the colour combinations. In order to write the partial widths in a more compact form, it is then very useful to define functions
\begin{eqnarray}
\pr(x,y) & = & |x|^2 + |y|^2 + \frac{2}{3} \, \RE x y^* \,, \notag \\
\pr(x,y,u,v) & = & xy^* + uv^* + \frac{1}{3} xv^* + \frac{1}{3} uy^* \,, 
\end{eqnarray}
which satisfy $\pr(x,x,y,y) = \pr(x,y)$, $\pr(x,x) = 8/3 |x|^2$. 
With this notation, the four-fermion contributions for $j,k\neq 3$ read
\begin{eqnarray}
\Gamma_\text{4F} & = & \frac{m_t}{2048\pi^3} \left[ \frac{m_t}{\Lambda} \right]^4 \notag \\
& & \times \left\{
\pr(\Cqq^{kji3},\Cqqp^{kji3}) + 4 \pr(\Cud^{i3kj},\Cudp^{i3kj})
+ \pr(\Cqup^{k3ij},\Cqu^{k3ij}) + \pr(\Cqdp^{ijk3},\Cqd^{ijk3}) \right. \notag \\
& & + \pr(\Cqqe^{i3kj},\Cqqep^{i3kj})
+ \pr(\Cqqe^{k3ij},\Cqqep^{k3ij}) + \pr(\Cqqe^{3ijk},\Cqqep^{3ijk})
+ \pr(\Cqqe^{ji3k},\Cqqep^{ji3k}) \notag \\
& & 
+ \left. \RE\! \left[ \pr(\Cqqe^{i3kj},\Cqqep^{k3ij},\Cqqep^{i3kj},\Cqqe^{k3ij})  
+ \pr(\Cqqe^{3ijk},\Cqqep^{ji3k},\Cqqep^{3ijk},\Cqqe^{ji3k})  \right] \right\} \,.
\label{ec:tdec-b}
\end{eqnarray}
When one of these indices equals three, substitutions in the $\bar L L \bar L L$ coefficients of Eq.~(\ref{ec:tdec-b}) may have to be performed because not all flavour combinations arise from independent operators, and some of them (in particular, the Hermitian ones) give a contribution twice larger. According to Tables~\ref{tab:res-bH} and \ref{tab:res-dH},
\begin{itemize}
\item for $k=3$ (decay $t \to b u_i \bar d_k$), $C_{qq^{(')}}^{kji3}$ must be replaced by $2 \, C_{qq^{(')}}^{3ii3}$ (which are real) if $i=j$, and by $C_{qq^{(')}}^{3ij3*}$ if $i>j$;
\item for $j=3$ (decay $t \to d_k u_i \bar b$), $C_{qq^{(')}}^{kji3}$ must be replaced
by $2 \, C_{qq^{(')}}^{i3i3}$ if $i=k$, and by $C_{qq^{(')}}^{i3k3}$ if $i>k$.
\end{itemize}
The interference between four-fermion operators and the SM amplitude is
\begin{eqnarray}
\Gamma_\text{int} & = & \frac{g^2 m_t}{512 \pi^3} \left[ \frac{m_t}{\Lambda} \right]^2
\left\{ \RE\! \left[ V_{3k} V_{ij}^* (\Cqqp^{kji3} + \ot \Cqq^{kji3}) \right]
( \Om_R^- + \Om_R^+ ) \right. \notag \\
& & \left. - \IM\! \left[ V_{3k} V_{ij}^* (\Cqqp^{kji3} + \ot \Cqq^{kji3}) \right]
 ( \Om_I^- + \Om_I^+ ) \right\} \,,
\label{ec:tdec-bint}
\end{eqnarray}
where the kinematic factors $\Om_{R,I}^\pm$ arise from the three-body phase space integration for invariant masses $m_{u_i \bar d_j} < M_W$ ($\Om_{R,I}^-$) and $m_{u_i \bar d_j} > M_W$ ($\Om_{R,I}^+$).
The first term between the curly brackets, proportional to the real part of coupling products, corresponds to the off-peak interference (in which the $W$ propagator is almost real), whereas the second term with the imaginary part is the peak contribution, where the $W$ propagator is imaginary. The corresponding phase space factors are
\begin{eqnarray}
\Om_R^- & \simeq & - \frac{9}{2} x_W^4 + \frac{11}{3} x_W^6 + (1-3 x_W^4 + 2 x_W^6) \log \frac{\Gamma_W}{M_W} + 3 \pi \frac{\Gamma_W}{m_t}(x_W^3-x_W^5) \,, \notag \\
\Om_R^+ & \simeq & -\frac{5}{6} - 2 x_W^2 + \frac{13}{2} x_W^4 - \frac{11}{3} x_W^6
+  (1-3 x_W^4 + 2 x_W^6) \log \frac{m_t^2-M_W^2}{\Gamma_W M_W} \notag \\
& & + 3 \pi \frac{\Gamma_W}{m_t}(x_W^3-x_W^5) \,, \notag \\
\Om_I^- & \simeq & \Om_I^+ \simeq \frac{\pi}{2}  (1-3 x_W^4 + 2 x_W^6) \,.
\end{eqnarray}
As it is expected, for the real part of the interference term there is a large cancellation in the total rate between the two phase space regions
$m_{u_i \bar d_j} < M_W$ and $m_{u_i \bar d_j} > M_W$,
in which the $W$ boson propagator changes sign.
For $m_t = 175$ GeV, $M_W = 80.4$ GeV, $\Gamma_W = 2.14$ GeV, we have $\Om_R^- = -3.367$, $\Om_R^+ = 3.385$ and the sum $\Om_R^- + \Om_R^+ = 0.018$ is 200 times smaller. On the other hand, the interference of the imaginary part is practically equal at both sides of the peak. For illustration, we show in Fig.~\ref{fig:mw} the normalised $u\bar d$ invariant mass distribution for the decay $t \to b u \bar d$ within the SM and with $C_{qq'}^{3113} = 10$, $\Lambda = 1$ TeV.%
It has been obtained numerically by implementing in the generator {\tt Protos}~\cite{AguilarSaavedra:2008gt} the four-fermion vertices. As a cross-check of our results, it has been verified that the numerical values of the interference and quadratic contributions coincide with the analytical ones in Eqs.~(\ref{ec:tdec-bint}) and (\ref{ec:tdec-b}). We observe that the leading four-fermion operator contributions, suppressed with respect to the SM one by the ratio
\begin{equation}
\eta_\text{dec}^\text{CC} \equiv \frac{3}{8 \pi^2} \left[ \frac{M_W}{\Lambda} \right]^2 \Omega_R^\pm \simeq 8.2 \times 10^{-4} \frac{1}{\Lambda^2} ~\text{TeV}^2 \,,
\label{ec:Rdec}
\end{equation}
are rather small even for relatively large value of the effective operator coefficients. Therefore, the presence of four-fermion operators with fields $t \bar d_k \bar u_i d_j$ can better be detected in single top production, which is discussed in section~\ref{sec:5.1}.

Finally, we point out that only $O_{qq}^{kji3}$ and $O_{qq'}^{kji3}$ interfere with the SM amplitude in the limit of vanishing $u_i$, $\bar d_j$ masses. (The same coefficient replacements indicated above have to be performed for specific values of indices.)
The colour flow for the amplitude with $O_{qq'}^{kji3}$ is the same as the SM one,
$t_a \to d_{ka} u_{ib} \bar d_{jb}$, and hence the interference takes place for all colour combinations; for $O_{qq}^{kji3}$ the flow is the other one and thus the $1/3$ factor multiplying its coefficient.

\subsection{$t \to d_k e_i^+ \nu_j$}
\label{sec:4.2}

This leptonic decay is much simpler than its hadronic counterpart in the previous subsection, because there is only one colour flow and fewer gauge-invariant operators with these fields. As a result, only 4 four-fermion terms contribute to the amplitude (see Tables~\ref{tab:res-bL} and \ref{tab:res-dL}). The SM and four-fermion contributions, as well as their interference, are
\begin{eqnarray}
\Gamma_\text{SM} & = & \frac{g^2 m_t}{576 \pi} \left[ \frac{m_t}{M_W} \right]^2 |V_{3k}|^2 \delta_{ij}
(1 - 3 x_W^4 + 2 x_W^6) \,, \notag \\[1mm]
\Gamma_\text{4F} & = & \frac{m_t}{6144\pi^3} \left[ \frac{m_t}{\Lambda} \right]^4
\left[ 4|\Clqp^{j3ki}|^2 + |\Cqde^{jik3}|^2 + |\Cqle^{kij3}|^2 + |\Clqe^{jik3}|^2
+ \RE \Cqle^{kij3} \Clqe^{jik3*} 
\right] \,, \notag \\[2mm]
\Gamma_\text{int} & = & \frac{g^2 m_t}{768 \pi^3} \left[ \frac{m_t}{\Lambda} \right]^2 \delta_{ij} \left\{
\RE [ V_{3k} \Clqp^{j3ki} ] ( \Om_R^- + \Om_R^+ )
- \IM [ V_{3k} \Clqp^{j3ki} ] ( \Om_I^- + \Om_I^+ )
\right\} \,.
\end{eqnarray}
For $k=3$ (decay $t \to b e_i^+ \nu_j$) the coefficients $\Clqp^{j3ki}$ must be replaced by $2 \, \Clqp^{i33i}$ (which are real) if $i=j$, and by $\Clqp^{i3kj*}$ if $i>j$.

The leading effects of four-fermion operators in this decay, namely the interference with the $O_{\ell q'}$ operators, are suppressed by $(M_W/\Lambda)^2$ and numerical factors with respect to the leading SM contribution. Still, this decay may be the only place to probe these operators because they do not contribute to single top production in hadron collisions. As we can see from Tables~\ref{tab:op2} and \ref{tab:op4}, $O_{\ell q'}$ cannot be probed in $e^+ e^-$ collisions either. On the other hand, for $k=1,2$, $O_{\ell q'}^{j3ki}$ also give $t \bar u_k \nu_i \bar \nu_j$ terms which can mediate a FCN decay $t \to u_k \bar \nu_i \nu_j$, discussed in section \ref{sec:4.4}.

\subsection{$t \to u_k  u_i \bar u_j$}
\label{sec:4.3}

This decay does not have a SM contribution but the calculation of the width is still non-trivial due to the presence of two up-type quarks in the final state. We have to distinguish the cases $i \neq k$ and $i=k$. In the first case there are 12 contributions to the amplitude, 6 corresponding to operators $(\bar u_i u_j) \, (\bar u_k t)$
and 6 to $(\bar u_k u_j) \, (\bar u_i t)$, with $i$ and $k$ interchanged. There are two colour flows $t_a \to u_{ka} u_{ib} \bar u_{jb}$ and $t_a \to u_{kb} u_{ia} \bar u_{jb}$ which only interfere for $a=b$. After averaging over colours, the resulting partial width can be compactly written as
\begin{eqnarray}
\Gamma_\text{4F} & = & \frac{m_t}{2048\pi^3} \left[ \frac{m_t}{\Lambda} \right]^4 \left[ \pr( \Cqq^{kji3}+\Cqqp^{ijk3}, \Cqq^{ijk3}+\Cqqp^{kji3})
+ \pr(\Cuu^{kji3},\Cuu^{ijk3}) 
 \right. \notag \\[1mm]
& & \left.
+ \pr(\Cqup^{k3ij},\Cqu^{k3ij}) + \pr(\Cqup^{i3kj}, \Cqu^{i3kj}) + \pr(\Cqup^{ijk3},\Cqu^{ijk3})
+ \pr(\Cqup^{kji3},\Cqu^{kji3})  \right] \,. \notag \\
\label{ec:t3u}
\end{eqnarray}
When $i=k$ there are 6 operators and only one colour flow, $t_a \to u_{ia} u_{ib} \bar u_{jb}$, because the two quarks $u_{ia}$, $u_{ib}$ are precisely distinguished by colour when $a \neq b$. Whereas, for $a = b$ they are identical particles and the amplitudes get two contributions from each Lagrangian term (with a $1/2$ symmetry factor). The final result, after colour averaging, is
\begin{eqnarray}
\Gamma_\text{4F} & = & \frac{m_t}{2048\pi^3} \left[ \frac{m_t}{\Lambda} \right]^4
\left[ \ft |\Cqq^{iji3}+\Cqqp^{iji3}|^2 + \ft |\Cuu^{iji3}|^2
+ \pr(\Cqup^{i3ij},\Cqu^{i3ij}) \right. \notag \\[1mm]
& & \left. + \pr(\Cqup^{iji3},\Cqu^{iji3}) \right] \,.
\end{eqnarray}
The latter expression (independently calculated) corresponds to Eq.~(\ref{ec:t3u}) setting $i=k$ and dividing by two (note that $\pr(x,x)=8/3|x|^2$). This relation can be easily understood from the previous considerations:
\begin{itemize}
\item Case $a \neq b$: for $i \neq k$ there are two sets of contributing operators which differ by the interchange $i \leftrightarrow k$, each set contributing to one of the possible colour flows. For $i = k$ there is only one set and only one colour flow. Then, setting $i=k$ in the partial width counts twice each contribution.
\item Case $a = b$: for $i \neq k$ there are two contributions from operators differing by the exchange of $i$ and $k$ while for $i = k$ each operator gives two terms in the amplitude because the two final state $u_i$ quarks are identical. However, in the latter case there is a symmetry factor of $1/2$.
\end{itemize}
It is worthwhile pointing out that this relation for the partial widths with $i \neq k$ and $i=k$ does not hold for the differential quantities, {\em i.e.} the angular distributions are not the same.

The possible effect of FCN four-fermion operators in top decays can be measured by the ratio of the prefactor in $\Gamma_\text{4F}$ over the SM width,
\begin{equation}
\eta_\text{dec}^\text{NC} \equiv
\frac{m_t/\Gamma_t}{2048 \pi^3} \left[ \frac{m_t}{\Lambda} \right]^4
\simeq 1.7 \times 10^{-6} \frac{1}{\Lambda^4} ~\text{TeV}^4 \,,
\label{ec:Rdec2}
\end{equation}
which determines the branching ratio for these decays, up to effective operator coefficients, once that the scale $\Lambda$ is set. The small value of this quantity implies that it is expected that new effects in FCN single top production (section~\ref{sec:5.2}) would be much easier to spot.

\subsection{$t \to u_k e_i^+ e_j^-$ and $t \to u_k \bar \nu_i \nu_j$}
\label{sec:4.4}

The computation of the decay rates for these processes, with a trivial colour structure and no SM contribution, is rather straightforward.
The coefficients of the eight operators contributing to $t \to u_k e_i^+ e_j^-$ are given in Table~\ref{tab:res-uL}. The partial width for this mode is
\begin{eqnarray}
\Gamma_\text{4F} & = & \frac{m_t}{6144\pi^3} \left[ \frac{m_t}{\Lambda} \right]^4
\left[ 4 |\Clq^{jik3}|^2 + 4 |\Ceu^{jik3}|^2 + |\Clu^{j3ki}|^2 + |\Cqe^{kij3}|^2
+ |\Cqle^{kij3}|^2 \right. \notag \\
& & \left.
+ |\Clqe^{jik3}|^2 + \RE \Cqle^{kij3} \Clqe^{jik3*} + |\Cqle^{3jik}|^2
+ |\Clqe^{ij3k}|^2 + \RE \Cqle^{3jik} \Clqe^{ij3k*}
\right] \,.
\end{eqnarray}
The decay $t \to u_k \bar \nu_i \nu_j$ is completely analogous but involving only two four-fermion terms, which can be read from Table~\ref{tab:res-tunn}. The corresponding partial width is
\begin{eqnarray}
\Gamma_\text{4F} & = & \frac{m_t}{6144\pi^3} \left[ \frac{m_t}{\Lambda} \right]^4
\left[ 4 |\Clq^{jik3}+\Clqp^{j3ki}|^2 + |\Clu^{j3ki}|^2 \right] \,.
\end{eqnarray}
These FCN decays are suppressed by $(m_t/\Lambda)^4$ as the previous decay $t \to u_k  u_i \bar u_j$ but, in contrast, the four-fermion terms involved do not contribute to single top production at hadron colliders. For $i=j=1$, $t \bar u_k e \bar e$ terms can be probed in single top production at an $e^+ e^-$ collider, but this is not the case for other lepton flavours, nor for terms involving two neutrinos.
\section{Single top production}
\label{sec:5}

After warming up with top decay width calculations, we present here the results for crossing symmetry-related processes: single top production in association with a light quark. Although the matrix elements are the same in both cases, in the latter the phase space integration introduces kinematical differences among effective operator contributions, and initial state parton distribution functions (PDFs) between processes. This makes the analysis much more cumbersome. We will discuss in turn charged current processes at LHC and Tevatron, FCN single top production at the same machines and single top production in $e^+ e^-$ collisions.

\subsection{Charged current processes in $pp$, $p \bar p$ collisions}
\label{sec:5.1}

These processes are the counterpart of the top decays studied in section~\ref{sec:4.1}. For a given initial and final state there are, in addition to a SM amplitude (possibly suppressed by small CKM matrix elements), 16 contributing four-fermion terms, 8 corresponding to each colour flow, with 4 vector and 4 scalar Lorentz structures. All the cross sections for these processes, namely
\begin{align}
& \sigma(u_i d_k \to d_j t) \,,\quad \sigma(\bar u_i \bar d_k \to \bar d_j \bar t) \,, \notag \\
& \sigma(\bar d_j d_k \to \bar u_i t) \,,\quad \sigma(d_j \bar d_k \to u_i \bar t) \,, \notag \\
& \sigma(u_i \bar d_j \to \bar d_k t) \,,\quad \sigma(\bar u_i d_j \to d_k \bar t)
\end{align}
can be written as
\begin{eqnarray}
\sigma & = & A_0 |V_{3k}|^2 |V_{ij}|^2 + \frac{A_\text{int}}{\Lambda^2} \, \RE\! \left[ V_{3k} V_{ij}^* (\Cqqp^{kji3} + \ot \Cqq^{kji3}) \right] \notag \\
& & + \frac{A_1}{\Lambda^4} \left[ \pr(\Cqq^{kji3},\Cqqp^{kji3}) + 4 \pr(\Cud^{i3kj},\Cudp^{i3kj}) \right] \notag \\
& & + \frac{A_2}{\Lambda^4} \left[  \pr(\Cqup^{k3ij},\Cqu^{k3ij}) + \pr(\Cqdp^{ijk3},\Cqd^{ijk3}) + \pr(\Cqqe^{k3ij},\Cqqep^{k3ij}) + \pr(\Cqqe^{ji3k},\Cqqep^{ji3k}) \right] \notag \\
& & + \frac{A_3}{\Lambda^4} \left[ \pr(\Cqqe^{i3kj},\Cqqep^{i3kj}) + \pr(\Cqqe^{3ijk},\Cqqep^{3ijk}) \right] \notag \\
& & + \frac{A_4}{\Lambda^4} \,
\RE\! \left[ \pr(\Cqqe^{i3kj},\Cqqep^{k3ij},\Cqqep^{i3kj},\Cqqe^{k3ij})  
+ \pr(\Cqqe^{3ijk},\Cqqep^{ji3k},\Cqqep^{3ijk},\Cqqe^{ji3k}) \right] \,,
\label{ec:cross-cc}
\end{eqnarray}
where the numerical factors $A_0$, $A_\text{int}$, $A_{1-4}$ depend on the specific process, as well as the collider ($pp$ or $p \bar p$) and CM energy.
Notice that $A_0$ times the appropriate CKM matrix elements in the first term on Eq.~(\ref{ec:cross-cc}) simply give the SM leading-order (LO) single top cross sections for $t$- and $s$-channel single top production subprocesses. In the above equation, the same replacements in operator coefficients done for top decays must be performed for specific index values:
\begin{itemize}
\item for $k=3$, $C_{qq^{(')}}^{kji3}$ must be replaced by $2 \, C_{qq^{(')}}^{3ii3}$ (which are real) if $i=j$, and by $C_{qq^{(')}}^{3ij3*}$ if $i>j$;
\item for $j=3$, $C_{qq^{(')}}^{kji3}$ must be replaced
by $2 \, C_{qq^{(')}}^{i3i3}$ if $i=k$, and by $C_{qq^{(')}}^{i3k3}$ if $i>k$.
\end{itemize}
The factors $A_i$ for LHC with 14 TeV are collected in Table~\ref{tab:cross-cc14}, for 7 TeV in Table~\ref{tab:cross-cc7} and for Tevatron in Table~\ref{tab:cross-cc2}. They have been computed using CTEQ6L1 PDFs~\cite{Pumplin:2002vw} with $Q=m_t$. There are more sophisticated factorisation scale choices for which SM single top LO cross sections are closer to next-to-leading order ones, but we prefer this simpler one, also bearing in mind that we are mainly interested in four-fermion contributions.

\begin{table}[p]
\begin{small}
\begin{center}
\begin{tabular}{lccccccc}
$u_i d_k \to d_j t$  & $A_0$ & $A_\text{int}$ & $A_1$ & $A_2,A_3$ & $A_4$ \\
\hline
$i=1$ \quad $k=1$ & 437   & -51.6  & 14.9  & 5.03  & -4.82 & $t$ \\
             & 55.3  & -5.33  & 0.530 & 0.185 & -0.161 & $\bar t$ \\
$i=2$ \quad $k=1$ & 81.7  & -8.10  & 0.915 & 0.317 & -0.281 & $t$ \\
             & 30.9  & -2.88  & 0.247 & 0.087 & -0.073 & $\bar t$ \\
$i=1$ \quad $k=2$ & 187   & -19.6  & 2.91  & 0.997 & -0.911 & $t$ \\ 
             & 36.3  & -3.42  & 0.310 & 0.109 & -0.093 & $\bar t$ \\
$i=2$ \quad $k=2$ & 19.5  & -1.78  & 0.142 & 0.050 & -0.042 & $t/\bar t$ \\
$i=1$ \quad $k=3$ & 106   & -10.7  & 1.27  & 0.438 & -0.390 & $t$ \\
             & 17.4  & -1.59  & 0.123 & 0.044 & -0.036 & $\bar t$   \\
$i=2$ \quad $k=3$ & 8.89  & -0.784 & 0.054 & 0.019 & -0.015 & $t/\bar t$
\end{tabular}
\vspace{5mm}

\begin{tabular}{lcccccc}
$\bar d_j d_k \to \bar u_i t$  & $A_0$ & $A_\text{int}$ & $A_1,A_2$ & $A_3,A_4$ \\
\hline
$j=1$ \quad $k=1$ & 120  & -9.93  & 0.783 & 2.28  & $t/\bar t$ \\
$j=2$ \quad $k=1$ & 85.7 & -6.86  & 0.475 & 1.37  & $t$ \\
             & 33.9 & -2.54  & 0.133 & 0.380 & $\bar t$ \\
$j=1$ \quad $k=2$ & 33.9 & -2.54  & 0.133 & 0.380 & $t$ \\
             & 85.7 & -6.86  & 0.475 & 1.37  & $\bar t$ \\
$j=2$ \quad $k=2$ & 21.7 & -1.60  & 0.078 & 0.221 & $t/\bar t$ \\
$j=1$ \quad $k=3$ & 16.4 & -1.19  & 0.054 & 0.154 & $t$ \\
             & 46.0 & -3.53  & 0.205 & 0.588 & $\bar t$ \\
$j=2$ \quad $k=3$ & 10.2 & -0.724 & 0.031 & 0.088 & $t/\bar t$
\end{tabular}
\vspace{5mm}

\begin{tabular}{lccccc}
$u_i \bar d_j \to \bar d_k t$ & $A_0$ & $A_\text{int}$ & $A_1,A_3$ & $A_2,A_4$ \\
\hline
$i=1$ \quad $j=1$ & 3.92  & 2.46  & 1.63  & 4.78  & $t$ \\
             & 2.24  & 1.28  & 0.65  & 1.89  & $\bar t$ \\
$i=1$ \quad $j=2$ & 3.07  & 1.81  & 1.00  & 2.90  & $t$ \\
             & 0.713 & 0.345 & 0.108 & 0.309 & $\bar t$ \\
$i=1$ \quad $j=3$ & 1.89  & 1.03  & 0.441 & 1.28  & $t$ \\
             & 0.363 & 0.164 & 0.044 & 0.124 & $\bar t$ \\
$i=2$ \quad $j=1$ & 0.615 & 0.292 & 0.087 & 0.247 & $t$ \\
             & 1.47  & 0.782 & 0.317 & 0.913 & $\bar t$ \\
$i=2$ \quad $j=2$ & 0.404 & 0.184 & 0.049 & 0.139 & $t/\bar t$ \\
$i=2$ \quad $j=3$ & 0.193 & 0.083 & 0.019 & 0.053 & $t/\bar t$
\end{tabular}
\end{center}
\end{small}
\caption{Numerical factors for single top cross sections at LHC with 14 TeV. The units of $A_0$, $A_\text{int}$ and $A_{1-4}$ are pb, $\text{pb} \cdot \text{TeV}^2$ and $\text{pb} \cdot \text{TeV}^4$, respectively. The labels $t$, $\bar t$ indicate whether the factors correspond to the processes in the left column or the charge conjugate.}
\label{tab:cross-cc14}
\end{table}

\begin{table}[p]
\begin{small}
\begin{center}
\begin{tabular}{lccccccc}
$u_i d_k \to d_j t$  & $A_0$ & $A_\text{int}$ & $A_1$ & $A_2,A_3$ & $A_4$ \\
\hline
$i=1$ \quad $k=1$ & 192   & -21.0  & 3.21    & 1.10    & -1.01    & $t$ \\
             & 14.0  & -1.24  & 0.0811  & 0.0290  & -0.0230  & $\bar t$ \\
$i=2$ \quad $k=1$ & 22.8  & -2.07  & 0.148   & 0.0527  & -0.0430  & $t$ \\
             & 6.86  & -0.586 & 0.0336  & 0.0121  & -0.0094  & $\bar t$ \\
$i=1$ \quad $k=2$ & 61.7  & -5.93  & 0.533   & 0.187   & -0.159   & $t$ \\ 
             & 8.43  & -0.731 & 0.0445  & 0.0160  & -0.0125  & $\bar t$ \\
$i=2$ \quad $k=2$ & 3.98  & -0.334 & 0.181   & 0.0660  & -0.00498 & $t/\bar t$ \\
$i=1$ \quad $k=3$ & 30.7  & -2.82  & 0.211   & 0.0749  & -0.0616  & $t$ \\
             & 3.51  & -0.294 & 0.0159  & 0.00580  & -0.00437 & $\bar t$   \\
$i=2$ \quad $k=3$ & 1.56  & -0.128 & 0.00624 & 0.00229 & -0.00167 & $t/\bar t$
\end{tabular}
\vspace{5mm}

\begin{tabular}{lcccccc}
$\bar d_j d_k \to \bar u_i t$  & $A_0$ & $A_\text{int}$ & $A_1,A_2$ & $A_3,A_4$ \\
\hline
$j=1$ \quad $k=1$ & 39.0 & -2.95  & 0.146  & 0.418  & $t/\bar t$ \\
$j=2$ \quad $k=1$ & 25.3 & -1.86  & 0.0844 & 0.239  & $t$ \\
             & 8.05 & -0.558 & 0.0205 & 0.0573 & $\bar t$ \\
$j=1$ \quad $k=2$ & 8.05 & -0.558 & 0.0205 & 0.0573 & $t$ \\
             & 25.3 & -1.86  & 0.0844 & 0.239  & $\bar t$ \\
$j=2$ \quad $k=2$ & 4.74 & -0.322 & 0.0112 & 0.0311 & $t/\bar t$ \\
$j=1$ \quad $k=3$ & 3.36 & -0.225 & 0.00746 & 0.0205 & $t$ \\
             & 11.8 & -0.835 & 0.0331  & 0.0928 & $\bar t$ \\
$j=2$ \quad $k=3$ & 1.92 & -0.127 & 0.00402 & 0.0110 & $t/\bar t$
\end{tabular}
\vspace{5mm}

\begin{tabular}{lccccc}
$u_i \bar d_j \to \bar d_k t$ & $A_0$ & $A_\text{int}$ & $A_1,A_3$ & $A_2,A_4$ \\
\hline
$i=1$ \quad $j=1$ & 1.60   & 0.876  & 0.319  & 0.917  & $t$ \\
             & 0.807  & 0.400  & 0.118  & 0.337  & $\bar t$ \\
$i=1$ \quad $j=2$ & 1.16   & 0.590  & 0.187  & 0.535  & $t$ \\
             & 0.187  & 0.0783 & 0.0159 & 0.0441 & $\bar t$ \\
$i=1$ \quad $j=3$ & 0.623  & 0.291  & 0.0753 & 0.213  & $t$ \\
             & 0.0827 & 0.0325 & 0.0058 & 0.0161 & $\bar t$ \\
$i=2$ \quad $j=1$ & 0.156  & 0.0635 & 0.0121 & 0.0334 & $t$ \\
             & 0.470  & 0.215  & 0.0528 & 0.149  & $\bar t$ \\
$i=2$ \quad $j=2$ & 0.0934 & 0.0368 & 0.0067 & 0.0184 & $t/\bar t$ \\
$i=2$ \quad $j=3$ & 0.0384 & 0.0143 & 0.0023 & 0.0062 & $t/\bar t$
\end{tabular}
\end{center}
\end{small}
\caption{Numerical factors for single top cross sections at LHC with 7 TeV. The units of $A_0$, $A_\text{int}$ and $A_{1-4}$ are pb, $\text{pb} \cdot \text{TeV}^2$ and $\text{pb} \cdot \text{TeV}^4$, respectively. The labels $t$, $\bar t$ indicate whether the factors correspond to the processes in the left column or the charge conjugate.}
\label{tab:cross-cc7}
\end{table}

\begin{table}[t]
\begin{small}
\begin{center}
\begin{tabular}{lccccccc}
$u_i d_k \to d_j t$  & $A_0$ & $A_\text{int}$ & $A_1$ & $A_2,A_3$ & $A_4$ \\
\hline
$i=1$ \quad $k=1$ & 4230  & -330  & 12.8  & 48.3  & -3.16  & $t/\bar t$ \\
$i=2$ \quad $k=1$ & 443   & -32.3 & 1.06  & 0.408 & -0.245 & $t/\bar t$ \\
$i=1$ \quad $k=2$ & 1800  & -138  & 5.22  & 1.97   & -1.27   & $t/\bar t$ \\ 
$i=2$ \quad $k=2$ & 59.5  & -4.09 & 0.117 & 0.046  & -0.025  & $t/\bar t$ \\
$i=1$ \quad $k=3$ & 599   & -44.4 & 1.53  & 0.583  & -0.359  & $t/\bar t$ \\
$i=2$ \quad $k=3$ & 16.2  & -1.09 & 0.030 & 0.012  & -0.0061 & $t/\bar t$
\end{tabular}
\vspace{5mm}

\begin{tabular}{lcccccc}
$\bar d_j d_k \to \bar u_i t$  & $A_0$ & $A_\text{int}$ & $A_1,A_2$ & $A_3,A_4$ \\
\hline
$j=1$ \quad $k=1$ & 4640 & -302  & 8.46  & 22.9  & $t/\bar t$ \\
$j=2$ \quad $k=1$ & 605  & -36.1 & 0.821 & 2.15  & $t/\bar t$ \\
$j=1$ \quad $k=2$ & 605  & -36.1 & 0.821 & 2.15  & $t/\bar t$ \\
$j=2$ \quad $k=2$ & 86.3 & -4.89 & 0.098 & 0.251 & $t/\bar t$ \\
$j=1$ \quad $k=3$ & 189  & -11.0 & 0.234 & 0.607 & $t/\bar t$ \\
$j=2$ \quad $k=3$ & 24.1 & -1.34 & 0.026 & 0.066 & $t/\bar t$
\end{tabular}
\vspace{5mm}

\begin{tabular}{lccccc}
$u_i \bar d_j \to \bar d_k t$ & $A_0$ & $A_\text{int}$ & $A_1,A_3$ & $A_2,A_4$ \\
\hline
$i=1$ \quad $j=1$ & 285   & 120   & 20.3  & 55.4  & $t/\bar t$ \\
$i=1$ \quad $j=2$ & 47.7  & 16.2  & 1.98  & 5.25  & $t/\bar t$ \\
$i=1$ \quad $j=3$ & 16.8  & 5.34  & 0.583 & 1.53  & $t/\bar t$ \\
$i=2$ \quad $j=1$ & 12.7  & 3.92  & 0.408 & 1.06  & $t/\bar t$ \\
$i=2$ \quad $j=2$ & 1.85  & 0.518 & 0.046 & 0.116 & $t/\bar t$ \\
$i=2$ \quad $j=3$ & 0.513 & 0.138 & 0.012 & 0.028 & $t/\bar t$
\end{tabular}
\end{center}
\end{small}
\caption{Numerical factors for single top cross sections at Tevatron. The units of $A_0$, $A_\text{int}$ and $A_{1-4}$ are fb, $\text{fb} \cdot \text{TeV}^2$ and $\text{fb} \cdot \text{TeV}^4$, respectively. The labels $t/\bar t$ indicate that the factors are equal for the processes in the left column and the charge conjugate.}
\label{tab:cross-cc2}
\end{table}

For $s$-channel production, it is remarkable to find that the quadratic contributions from some four-fermion operators which do not interfere with the SM amplitude can be as large as those from the interference terms. For example, setting $i=j=1$, $k=3$ and neglecting CKM mixing for simplicity we have
\begin{eqnarray}
\sigma_\text{int}(u \bar d \to \bar b t) & = & \frac{4.92}{\Lambda^2} \,\RE\! \left[ \Cqqp^{3113} + \ot \Cqq^{3113} \right] ~\text{pb}\cdot\text{TeV}^2 \,, \notag \\[1mm]
\sigma_\text{4F}(u \bar d \to \bar b t) & = & \frac{4.78}{\Lambda^4} \left[ |\Cqup^{3311}|^2 + |\Cqu^{3311}|^2 + \twt \RE \Cqup^{3311} \Cqu^{3311*} \right] ~\text{pb} \cdot \text{TeV}^4 + \dots \,,\notag \\
\end{eqnarray}
where we have omitted quadratic contributions from other operators in the second equation. For $s$-channel production the quadratic term is large because it is not suppressed by the $s$-channel $W$ propagator as the linear and SM terms are. In contrast, for the $t$-channel processes $u b \to b t$ and $\bar d b \to \bar u t$ the interference terms ($A_\text{int}$) are a factor of five larger than the quadratic ones ($A_{1-4}$).

It is also worth comparing the four-fermion operator effects in top decays and
single top production. In the former processes, the leading corrections are proportional to the small ratio $\eta_\text{dec}^\text{CC}$ in Eq.~(\ref{ec:Rdec}), of order $10^{-3}$. On the other hand, for single top production the leading effects, relative to the SM cross sections, are proportional to
\begin{equation}
\eta_\text{prod}^\text{CC} \equiv \frac{A_\text{int}/\Lambda^2}{A_0} \,.
\end{equation}
For example, for $u b \to d t$ at LHC (with 14 and 7 TeV) this ratio is around $\eta_\text{prod}^\text{CC} = 0.1/\Lambda^2~\text{TeV}^2$, 100 times larger than $\eta_\text{dec}^\text{CC}$. This means that it will be difficult, using bounds from other processes, to get rid of possible four-fermion operator contributions to $t$-channel single top production to obtain a model-independent measurement of $V_{tb}$~\cite{AguilarSaavedra:2010wf}.

Besides, we point out that can recover previous results~\cite{Cao:2007ea} by setting $\Cqqp^{3ij3} \equiv 2 (\Lambda^2/v^2) V_{ij} \, \mathcal{G}_{4f}$ (with $\mathcal{G}_{4f}$ a real parameter). Summing all contributions from the different sub-processes in Table~\ref{tab:cross-cc14} we get the inclusive $t$- and $s$-channel cross sections for LHC at 14 TeV
\begin{eqnarray}
\sigma_t & = & \sigma_t^0 (1-2.95 \, \mathcal{G}_{4f} + \dots) \,, \notag \\
\sigma_s & = & \sigma_s^0 (1+19.43 \, \mathcal{G}_{4f} + \dots) \,,
\end{eqnarray}
where $\sigma_{t,s}^0$ are the SM cross sections and the dots stand for quadratic terms which, as we have found, can be of the same size as the linear ones for $s$-channel production. These equations agree very well with Ref.~\cite{Cao:2007ea}, and the small differences (-3.5\% and -1.1\%, respectively) in the coefficients of $\mathcal{G}_{4f}$ can be attributed to a different choice of PDFs or factorisation scale.

\subsection{Flavour-changing neutral processes in $pp$, $p \bar p$ collisions}
\label{sec:5.2}

There are several processes of FCN single top production in hadron collisions, absent in the SM, to which four-fermion operators can contribute,
\begin{align}
& u_i u_k \to u_j t \,, \notag \\
& \bar u_i \bar u_k \to \bar u_j \bar t \,, \notag \\
& u_i \bar u_j \to u_k t \,.
\end{align}
They are related by crossing symmetry and/or charge conjugation to the FCN top decays studied in section~\ref{sec:4.2}, and thus have the same matrix elements, with 12 four-fermion terms (from 14 operators) contributing to the amplitudes for $i \neq k$ and 6 terms (from 7 operators) for $i=k$. In the former case, the cross sections can be written as
\begin{eqnarray}
\sigma & = & \frac{B_1}{\Lambda^4} \left[ \pr(\Cqq^{kji3}+\Cqqp^{ijk3},\Cqq^{ijk3}+\Cqqp^{kji3}) + \pr(\Cuu^{kji3},\Cuu^{ijk3}) \right] \notag \\
& & + \frac{B_2}{\Lambda^4} \left[  \pr(\Cqup^{k3ij},\Cqu^{k3ij}) + \pr(\Cqup^{ijk3},\Cqu^{ijk3})  \right] \notag \\
& & + \frac{B_3}{\Lambda^4} \left[ \pr(\Cqup^{i3kj},\Cqu^{i3kj}) + \pr(\Cqup^{kji3},\Cqu^{kji3}) \right] \,.
\end{eqnarray}
These equations are also valid for $i=k$, with a $1/2$ symmetry factor for $u_i u_k \to u_j t$ and $\bar u_i \bar u_k \to \bar u_j \bar t$ absorbed into the definition of the corresponding $B_i$ coefficients (see the discussion in section~\ref{sec:4.3} regarding the relation between $i \neq k$ and $i=k$). Alternatively, one can also use a simpler expression obtained from the one above by setting $i=k$, 
\begin{eqnarray}
\sigma & = & \frac{B_1}{\Lambda^4} \left[ \et |\Cqq^{iji3}+\Cqqp^{iji3}|^2 + \et |\Cuu^{iji3}|^2 \right] \notag \\
& & + \frac{B_2+B_3}{\Lambda^4} \left[  \pr(\Cqup^{i3ij},\Cqu^{i3ij}) + \pr(\Cqup^{iji3},\Cqu^{iji3})  \right] \,.
\end{eqnarray}
The factors $B_i$ for LHC with 14 TeV are given in Table~\ref{tab:cross-nc14}, for 7 TeV in Table~\ref{tab:cross-nc7} and for Tevatron in Table~\ref{tab:cross-nc2}. They are computed using CTEQ6L1 PDFs~\cite{Pumplin:2002vw} with $Q=m_t$. These FCN processes have been previously considered in Ref.~\cite{Ferreira:2006xe} but unfortunately the cross sections provided are totally inclusive, summing charged current processes from several flavours as well, and a direct comparison with their results is difficult.

\begin{table}[htb]
\begin{small}
\begin{center}
\begin{tabular}{lcccc}
$u_i u_k \to u_j t$  & $B_1$ & $B_2,B_3$ \\
\hline
$i=1$ \quad $k=1$ & 14.9   & 5.03   & $t$ \\
             & 0.214  & 0.0747 & $\bar t$ \\
$i=1$ \quad $k=2$ & 1.95   & 0.674  & $t$ \\
             & 0.195  & 0.0690 & $\bar t$ \\
$i=2$ \quad $k=2$ & 0.0431 & 0.0153 & $t/\bar t$
\end{tabular}
\quad
\begin{tabular}{lcccc}
$u_i \bar u_j \to u_k t$  & $B_1,B_3$ & $B_2$ \\
\hline
$i=1$ \quad $j=1$  & 1.35   & 3.95   & $t/\bar t$ \\
$i=1$ \quad $j=2$  & 0.674  & 1.95   & $t$ \\
              & 0.0690 & 0.195  & $\bar t$ \\
$i=2$ \quad $j=1$  & 0.0690 & 0.195  & $t$ \\
              & 0.674  & 1.95   & $\bar t$ \\
$i=2$ \quad $j=2$  & 0.0307 & 0.0864 & $t/\bar t$ 
\end{tabular}
\end{center}
\end{small}
\caption{Numerical factors for single top cross sections at LHC with 14 TeV. The units of $B_{1-3}$ are $\text{pb} \cdot \text{TeV}^4$. The labels $t$, $\bar t$ indicate whether the factors correspond to the processes in the left column or the charge conjugate.}
\label{tab:cross-nc14}
\end{table}

\begin{table}[htb]
\begin{small}
\begin{center}
\begin{tabular}{lcccc}
$u_i u_k \to u_j t$  & $B_1$ & $B_2,B_3$ \\
\hline
$i=1$ \quad $k=1$ & 3.33    & 1.14    & $t$ \\
             & 0.0316  & 0.0114  & $\bar t$ \\
$i=1$ \quad $k=2$ & 0.337   & 0.119   & $t$ \\
             & 0.0260  & 0.00942 & $\bar t$ \\
$i=2$ \quad $k=2$ & 0.00517 & 0.00189 & $t/\bar t$
\end{tabular}
\quad
\begin{tabular}{lcccc}
$u_i \bar u_j \to u_k t$  & $B_1,B_3$ & $B_2$ \\
\hline
$i=1$ \quad $j=1$  & 0.259   & 0.740  & $t/\bar t$ \\
$i=1$ \quad $j=2$  & 0.119   & 0.337  & $t$ \\
              & 0.00942 & 0.0260 & $\bar t$ \\
$i=2$ \quad $j=1$  & 0.00942 & 0.0260 & $t$ \\
              & 0.119   & 0.337  & $\bar t$ \\
$i=2$ \quad $j=2$  & 0.00380 & 0.0104 & $t/\bar t$ 
\end{tabular}
\end{center}
\end{small}
\caption{Numerical factors for single top cross sections at LHC with 7 TeV. The units of $B_{1-3}$ are $\text{pb} \cdot \text{TeV}^4$. The labels $t$, $\bar t$ indicate whether the factors correspond to the processes in the left column or the charge conjugate.}
\label{tab:cross-nc7}
\end{table}

\begin{table}[htb]
\begin{small}
\begin{center}
\begin{tabular}{lcccc}
$u_i u_k \to u_j t$  & $B_1$ & $B_2,B_3$ \\
\hline
$i=1$ \quad $k=1$ & 7.49   & 2.81   & $t/\bar t$ \\
$i=1$ \quad $k=2$ & 2.64   & 1.01   & $t/\bar t$ \\
$i=2$ \quad $k=2$ & 0.0262 & 0.0104 & $t/\bar t$
\end{tabular}
\quad
\begin{tabular}{lcccc}
$u_i \bar u_j \to u_k t$  & $B_1,B_3$ & $B_2$ \\
\hline
$i=1$ \quad $j=1$  & 48.0   & 132    & $t/\bar t$ \\
$i=1$ \quad $j=2$  & 1.01   & 2.64   & $t/\bar t$ \\
$i=2$ \quad $j=1$  & 1.01   & 2.64   & $t/\bar t$  \\
$i=2$ \quad $j=2$  & 0.0210 & 0.0530 & $t/\bar t$ 
\end{tabular}
\end{center}
\end{small}
\caption{Numerical factors for single top cross sections at Tevatron. The units of $B_{1-3}$ are $\text{fb} \cdot \text{TeV}^4$. The labels $t/\bar t$ indicate that the factors are equal for the processes in the left column and the charge conjugate.}
\label{tab:cross-nc2}
\end{table}

The relative size of FCN single top production with respect to SM processes can be appreciated by calculating the ratios
\begin{equation}
\eta_\text{prod}^\text{NC} \equiv
\frac{B/\Lambda^4}{\sigma_\text{SM}}
\end{equation}
of the FCN cross sections (up to effective operator coefficients) over the SM single top cross section. For $u u \to u t$ this ratio takes values up to $\eta_\text{prod}^\text{NC} = 0.067/\Lambda^4 ~\text{TeV}^4$, which is $4 \times 10^4$ times larger than the corresponding quantity $\eta_\text{dec}^\text{NC}$ in the top decay, making very interesting the study of four-fermion operator effects in single top production.

\subsection{Single top production in $e^+ e^-$ collisions}

Single top production in $e^+ e^-$ collisions constitutes the best process to probe $t \bar u_k e \bar e$ four-fermion terms, if a high-energy International Linear Collider (ILC) is built.
The cross section for $e^+ e^- \to \bar u_k t$ with longitudinally polarised beams is
\begin{eqnarray}
\sigma(e_R^+ e_L^-) & = & \frac{s}{8\pi \Lambda^4} \frac{\beta^2}{(1+\beta)^3}
\left[ 4 |\Clq^{11k3}|^2 + |\Clu^{13k1}|^2 \right] (3+\beta) \,, \notag \\
\sigma(e_L^+ e_R^-) & = & \frac{s}{8\pi \Lambda^4} \frac{\beta^2}{(1+\beta)^3}
\left[ 4 |\Ceu^{11k3}|^2 + |\Cqe^{13k1}|^2 \right] (3+\beta) \,, \notag \\
\sigma(e_L^+ e_L^-) & = & \frac{s}{8\pi \Lambda^4} \frac{\beta^2}{(1+\beta)^3}
\left[ |\Cqle^{k113}|^2 (3+\beta) + 6 |\Clqe^{11k3}|^2 (1+\beta) \right. \notag \\
& & \left. + 6 \, \RE \Cqle^{k113} \Clqe^{11k3*} (1+\beta) \right] \,, \notag \\
\sigma(e_R^+ e_R^-) & = & \frac{s}{8\pi \Lambda^4} \frac{\beta^2}{(1+\beta)^3}
\left[ |\Cqle^{311k}|^2 (3+\beta) + 6 |\Clqe^{113k}|^2 (1+\beta) \right. \notag \\
& & \left. + 6 \, \RE \Cqle^{311k} \Clqe^{113k*} (1+\beta) \right] \,,
\end{eqnarray}
being $\sqrt s$ the CM energy and $\beta = (s-m_t^2)/(s+m_t^2)$ the top velocity in the CM frame. Note that for $e^+$ the subindex indicates the helicity, not the chirality. For $u_k \bar t$ production the cross sections are the same. Our expressions for the vector terms (first two equations) agree with those in Ref.~\cite{BarShalom:1999iy}, as it can be seen by translating our notation,
$V_{LL} \equiv \Clq^{11k3*}$, $V_{RR} \equiv \Ceu^{11k3*}$,
$V_{LR} \equiv -\Clu^{13k1*}/2$, $V_{RL} \equiv -\Cqe^{k113*}/2$.\footnote{Notice a missing factor of $1/2$ in Eq.~(32) of Ref.~\cite{BarShalom:1999iy}, because the total cross section is the average of the polarised ones.}
For the scalar terms, our operators $O_{q\ell\epsilon}^{311k}$ and $O_{\ell q\epsilon}^{113k}$ are equivalent to $S_{RR}$ and $T_{RR}$ in that reference (the last one obtained by a Fierz transformation of a scalar term) while our operators
$O_{q\ell\epsilon}^{k113}$ and $O_{\ell q \epsilon}^{11k3}$ were not included. As it can be found from Table~\ref{tab:res-uL},
these operators generate terms $(\bar u_{Lk} \, e_R ) (\bar e_L \, t_R )$ and
$(\bar u_{Lk} \, t_R) (\bar e_L \, e_R)$, respectively, plus the Hermitian conjugate. In the notation of Ref.~\cite{BarShalom:1999iy}, they would correspond to $S_{LL}$ terms.

\section{Top pair production}
\label{sec:6} 

For top pair production in hadron and $e^+ e^-$ collisions the multiplicity of sub-processes is much smaller than for single top production. However, the matrix elements (and thus the total cross sections) are complicated by the presence of more interference terms, proportional to $m_t^2$, which are not present in processes with three light quarks. We will first study $t \bar t$ production at LHC and Tevatron. The conspicuous process of like-sign top pair production in hadron collisions will be discussed in detail next. Finally, we will turn our attention to $t \bar t$ production in $e^+ e^-$ collisions.

\subsection{$t \bar t$ production in $pp$, $p \bar p$ collisions}
\label{sec:6.1}

We consider here the processes $\bar u_i u_j \to t \bar t$, $\bar d_i d_j \to t \bar t$, with $i,j=1,2$, for which there is a SM QCD contribution when $i=j$. (We do not include electroweak $t \bar t$ production in our calculations.) For $\bar u_i u_j$ there are, in addition, 12 independent four-fermion terms resulting from 14 effective operators, 6 terms for the colour flow $\bar u_{ib} u_{ja} \to t_a \bar t_b$ and 6 for $\bar u_{ia} u_{ja} \to t_b \bar t_b$.  For $\bar d_i d_j$ there are 16 independent four-fermion terms, 8 for each colour flow. We have checked that our matrix elements coincide with Ref.~\cite{Jung:2009pi}, for the subset of operators considered there.
For $u_i \bar u_i \to t \bar t$ the SM plus four-fermion contributions are
\begin{eqnarray}
\sigma(u_i \bar u_i) & = & D_0  + \frac{D_\text{int}}{\Lambda^2} \left[
\Cqqp^{ii33} + \Cqq^{3ii3} + \Cuu^{3ii3} - \Cqu^{i33i} - \Cqu^{3ii3} \right] \notag \\
& & + \frac{D_1}{\Lambda^4} \left[ \pr(\Cqq^{ii33}+\Cqqp^{3ii3},\Cqqp^{ii33}+\Cqq^{3ii3})
+ \pr(\Cuu^{ii33},\Cuu^{3ii3}) \right. \notag \\
& & \left.
+ \pr(\Cqup^{i33i},\Cqu^{i33i}) + \pr(\Cqup^{3ii3},\Cqu^{3ii3}) \right]  + \frac{D_2}{\Lambda^4} \, \pr(\Cqup^{33ii},\Cqu^{33ii}) \notag \\
& & + \frac{D_3}{\Lambda^4} \left[
\pr(\Cqq^{ii33}+\Cqqp^{3ii3},\Cqup^{i33i},\Cqqp^{ii33}+\Cqq^{3ii3},\Cqu^{i33i}) \right. \notag \\
&& \left. + \pr(\Cqup^{3ii3},\Cuu^{ii33},\Cqu^{3ii3},\Cuu^{3ii3})
\right] \,,
\end{eqnarray}
and for flavour-nondiagonal processes the four-fermion cross sections are
\begin{eqnarray}
\sigma(u \bar c,c \bar u) & = & \frac{D_1}{\Lambda^4} \left[ \pr(\Cqq^{1233}+\Cqqp^{3213},\Cqqp^{1233}+\Cqq^{3213})
+ \pr(\Cuu^{1233},\Cuu^{3213}) \right. \notag \\
& & \left.
+ \pr(\Cqup^{1332},\Cqu^{1332}) + \pr(\Cqup^{3213},\Cqu^{3213}) \right] \notag \\
& & + \frac{D_2}{\Lambda^4} \left[ \pr(\Cqup^{3312},\Cqu^{3312}) + \pr(\Cqup^{3321},\Cqu^{3321}) \right] \notag \\
& & + \frac{D_3}{\Lambda^4} \,\RE\! \left[
\pr(\Cqq^{1233}+\Cqqp^{3213},\Cqup^{1332},\Cqqp^{1233}+\Cqq^{3213},\Cqu^{1332}) \right. \notag \\
&& \left. + \pr(\Cqup^{3213},\Cuu^{1233},\Cqu^{3213},\Cuu^{3213})
\right] \,,
\end{eqnarray}
The numerical coefficients $D_i$ are collected in Table~\ref{tab:cross-tt14} for LHC at 14 TeV, in Table~\ref{tab:cross-tt7} for the same collider at 7 TeV and in Table~\ref{tab:cross-tt2} for Tevatron. We have used CTEQ6L1 PDFs with a factorisation scale $Q=m_t$. Notice that, except for different PDFs, the cross sections for $u \bar c$ and $c \bar u$ are equal, involving the same set of operator coefficients.

It is remarkable that at LHC with 14 TeV the quadratic terms multiplying $D_2$, corresponding to four-fermion terms which do not interfere with the SM, give a contribution which can be comparable to the interferences of other operators. For example, 
\begin{eqnarray}
\sigma_\text{int}(u \bar u) & = & \frac{9.04}{\Lambda^2} \left[
\Cqqp^{1133} + \Cqq^{3113} + \Cuu^{3113} - \Cqu^{1331} - \Cqu^{3113} \right] ~\text{pb} \cdot \text{TeV}^2 \,, \notag \\[1mm]
\sigma_\text{4F}(u \bar u) & = & \frac{6.82}{\Lambda^4} \left[ |\Cqup^{3311}|^2 + |\Cqu^{3311}|^2 + \twt \, \RE \Cqup^{3311} \Cqu^{3311*} \right] ~\text{pb} \cdot \text{TeV}^4 + \dots \,,
\end{eqnarray}
where we have omitted other contributions in the second equation.
Therefore, these operators are worth being investigated in detail, as well as the ones interfering with the SM.
For $d_i \bar d_i \to t \bar t$ the SM plus four-fermion contributions are
\begin{eqnarray}
\sigma(d_i \bar d_i) & = & D_0 + \frac{D_\text{int}}{\Lambda^2} \left[ \Cqqp^{ii33}+2\,\Cudp^{33ii} -\Cqu^{i33i}-\Cqd^{3ii3} \right] \notag \\
& & + \frac{D_1}{\Lambda^4} \left[ \pr(\Cqq^{ii33},\Cqqp^{ii33}) + 4 \pr(\Cud^{33ii},\Cudp^{33ii}) + \pr(\Cqup^{i33i},\Cqu^{i33i}) \right. \notag \\
& & \left. + \pr(\Cqdp^{3ii3},\Cqd^{3ii3}) + \oh \pr(\Cqqep^{i33i},\Cqqe^{i33i}) \right] \notag \\
& & + \frac{D_2}{\Lambda^4} \left[ \pr(\Cqqe^{33ii},\Cqqep^{33ii}) + \RE \pr(\Cqqe^{33ii},\Cqqep^{i33i},\Cqqep^{33ii},\Cqqe^{i33i})
\right] \notag \\
& & + \frac{D_3}{\Lambda^4} \left[ \pr(\Cqq^{ii33},\Cqup^{i33i},\Cqqp^{ii33},\Cqu^{i33i})+2 \pr(\Cqdp^{3ii3},\Cud^{33ii},\Cqd^{3ii3},\Cudp^{33ii})
\right] \,,
\end{eqnarray}
and for flavour non-diagonal combinations we have
\begin{eqnarray}
\sigma(d \bar s, s \bar d) & = & \frac{D_1}{\Lambda^4} \left[ \pr(\Cqq^{1233},\Cqqp^{1233}) + 4 \pr(\Cud^{3312},\Cudp^{3312}) + \pr(\Cqup^{1332},\Cqu^{1332}) \right. \notag \\
& & \left. + \pr(\Cqdp^{3213},\Cqd^{3213}) + \pr(\Cqqep^{1332},\Cqqe^{1332}) + \pr(\Cqqep^{2331},\Cqqe^{2331}) \right] \notag \\
& & + \frac{D_2}{\Lambda^4} \left[ \pr(\Cqqe^{3312},\Cqqep^{3312}) + \pr(\Cqqe^{3321},\Cqqep^{3321}) + \RE \pr(\Cqqe^{3312},\Cqqep^{1332},\Cqqep^{3312},\Cqqe^{1332})
\right. \notag \\
& & \left. 
+ \RE \pr(\Cqqe^{3321},\Cqqep^{2331},\Cqqep^{3321},\Cqqe^{2331}) \right] \notag \\
& & + \frac{D_3}{\Lambda^4} \,\RE\! \left[ \pr(\Cqq^{1233},\Cqup^{1332},\Cqqp^{1233},\Cqu^{1332})+2 \pr(\Cqdp^{3213},\Cud^{3312},\Cqd^{3213},\Cudp^{3312})
\right] \,. \notag \\
\end{eqnarray}

\begin{table}[htb]
\begin{small}
\begin{center}
\begin{tabular}{lcccccc}
$\bar u_i u_j \to t \bar t$  & $D_0$ & $D_\text{int}$ & $D_1$ & $D_2$ & $D_3$ \\
\hline
$i=1$ \quad $j=1$ & 42.5 & 9.04  & 4.72   & 6.82   & -1.07 \\
$i=1$ \quad $j=2$ & --   & --    & 0.0494 & 0.138  & -0.0209 \\
$i=2$ \quad $j=1$ & --   & --    & 0.564  & 1.61   & -0.158 \\
$i=2$ \quad $j=2$ & 2.11 & 0.348 & 0.0820 & 0.113 & -0.0393
\end{tabular}
\vspace{5mm}

\begin{tabular}{lcccccc}
$\bar d_i d_j \to t \bar t$  & $D_0$ & $D_\text{int}$ & $D_1$ & $D_2$ & $D_3$ \\
\hline
$i=1$ \quad $j=1$ & 26.4 & 5.51  & 2.70   & 3.88   & -0.648 \\
$i=1$ \quad $j=2$ & --   & --    & 0.103  & 0.290  & -0.0377 \\
$i=2$ \quad $j=1$ & --   & --    & 0.399  & 1.14   & -0.110 \\
$i=2$ \quad $j=2$ & 4.56 & 0.802 & 0.230  & 0.322 & -0.0916
\end{tabular}
\end{center}
\end{small}
\caption{Numerical factors for $t \bar t$ cross sections at LHC with 14 TeV. The units of $D_0$, $D_\text{int}$ and $D_{1-3}$ are pb, $\text{pb} \cdot \text{TeV}^2$ and $\text{pb} \cdot \text{TeV}^4$, respectively.}
\label{tab:cross-tt14}
\end{table}

\begin{table}[htb]
\begin{small}
\begin{center}
\begin{tabular}{lcccccc}
$\bar u_i u_j \to t \bar t$  & $D_0$ & $D_\text{int}$ & $D_1$ & $D_2$ & $D_3$ \\
\hline
$i=1$ \quad $j=1$ & 14.6  & 2.69   & 0.796   & 1.11    & -0.309 \\
$i=1$ \quad $j=2$ & --    & --     & 0.00532 & 0.0143  & -0.00329 \\
$i=2$ \quad $j=1$ & --    & --     & 0.00847 & 0.234   & -0.0387 \\
$i=2$ \quad $j=2$ & 0.323 & 0.0477 & 0.00780 & 0.104   & -0.00524
\end{tabular}
\vspace{5mm}

\begin{tabular}{lcccccc}
$\bar d_i d_j \to t \bar t$  & $D_0$ & $D_\text{int}$ & $D_1$ & $D_2$ & $D_3$ \\
\hline
$i=1$ \quad $j=1$ & 8.70  & 1.58  & 0.440   & 0.614  & -0.181 \\
$i=1$ \quad $j=2$ & --    & --    & 0.0127  & 0.0344 & -0.00713 \\
$i=2$ \quad $j=1$ & --    & --    & 0.0609  & 0.169  & -0.0272 \\
$i=2$ \quad $j=2$ & 0.903 & 0.141 & 0.0270  & 0.0364 & -0.0157
\end{tabular}
\end{center}
\end{small}
\caption{Numerical factors for $t \bar t$ cross sections at LHC with 7 TeV. The units of $D_0$, $D_\text{int}$ and $D_{1-3}$ are pb, $\text{pb} \cdot \text{TeV}^2$ and $\text{pb} \cdot \text{TeV}^4$, respectively.}
\label{tab:cross-tt7}
\end{table}

\begin{table}[htb]
\begin{small}
\begin{center}
\begin{tabular}{lcccccc}
$\bar u_i u_j \to t \bar t$  & $D_0$ & $D_\text{int}$ & $D_1$ & $D_2$ & $D_3$ \\
\hline
$i=1$ \quad $j=1$ & 4780 & 717   & 110     & 145    & -79.2 \\
$i=1$ \quad $j=2$ & --   & --    & 0.295   & 0.735  & -0.292 \\
$i=2$ \quad $j=1$ & --   & --    & 0.295   & 0.735  & -0.292 \\
$i=2$ \quad $j=2$ & 1.18 & 0.137 & 0.0123  & 0.0149 & -0.0142
\end{tabular}
\vspace{5mm}

\begin{tabular}{lcccccc}
$\bar d_i d_j \to t \bar t$  & $D_0$ & $D_\text{int}$ & $D_1$ & $D_2$ & $D_3$ \\
\hline
$i=1$ \quad $j=1$ & 868  & 120   & 15.3   & 19.7  & -12.9 \\
$i=1$ \quad $j=2$ & --   & --    & 0.235  & 0.587 & -0.238 \\
$i=2$ \quad $j=1$ & --   & --    & 0.235  & 0.587 & -0.238 \\
$i=2$ \quad $j=2$ & 6.82 & 0.806 & 0.0752 & 0.0916 & -0.0836
\end{tabular}
\end{center}
\end{small}
\caption{Numerical factors for $t \bar t$ cross sections at Tevatron. The units of $D_0$, $D_\text{int}$ and $D_{1-3}$ are fb, $\text{fb} \cdot \text{TeV}^2$ and $\text{fb} \cdot \text{TeV}^4$, respectively.}
\label{tab:cross-tt2}
\end{table}
The coefficients $D_i$ for these processes can be found in Tables~\ref{tab:cross-tt14}, \ref{tab:cross-tt7} and \ref{tab:cross-tt2} as well.
We point out that again there are effective operators which do not interfere with the SM but can give quadratic contributions of the same order of the interference terms at a CM energy of 14 TeV. For example,
\begin{eqnarray}
\sigma_\text{int}(d \bar d) & = & \frac{5.51}{\Lambda^2} \, \left[
\Cqqp^{1133}+2\Cudp^{3311} -\Cqu^{1331}-\Cqd^{3113} \right] ~\text{pb} \cdot \text{TeV}^2 \,, \notag \\[1mm]
\sigma_\text{4F}(d \bar d) & = & \frac{3.88}{\Lambda^4} \left[
|\Cqqep^{3311}|^2 + |\Cqqe^{3311}|^2 + \twt\, \RE \Cqqep^{3311} \Cqqe^{3311*}
\right] ~\text{pb} \cdot \text{TeV}^4 + \dots \,,
\end{eqnarray}
with additional terms omitted in the second equation.

For Tevatron, it is also of interest to provide expressions for the FB asymmetry of the top quark, motivated by an apparent disagreement between the CDF~\cite{Aaltonen:2008hc} and D0~\cite{:2007qb} measurements and the SM prediction. Let $\theta$ be the angle between the top quark momentum in the $t \bar t$ rest frame and the incoming proton momentum. 
The FB asymmetry is
\begin{equation}
A_\text{FB} = \frac{\sigma(\cos \theta > 0) - \sigma(\cos \theta < 0)}{\sigma(\cos \theta > 0) + \sigma(\cos \theta < 0)} \,.
\label{ec:afb}
\end{equation}
If this asymmetry is originated by new heavy physics contributing to $u \bar u \to t \bar t$, $d \bar d \to t \bar t$, it can be parameterised in terms of gauge-invariant effective operators. In order to calculate this asymmetry in the presence of four-fermion operators, we give here the forward and backward cross sections
\begin{equation}
\sigma^F \equiv \sigma(\cos \theta > 0) \,, \quad 
\sigma^B \equiv \sigma(\cos \theta < 0) \,,
\end{equation}
for $u \bar u \to t \bar t$, $d \bar d \to t \bar t$ at the tree level. For the former process they are
\begin{eqnarray}
\sigma^{F,B}(u \bar u) & = & 2.39~\text{pb}
+ \frac{D_\text{int}^{F,B}}{\Lambda^2}
\left[ \Cqqp^{1133} + \Cqq^{3113} + \Cuu^{3113} \right]
- \frac{\tilde D_\text{int}^{F,B}}{\Lambda^2}
\left[ \Cqu^{1331} + \Cqu^{3113}  \right] \notag \\
& & + \frac{D_1^{F,B}}{\Lambda^4}
\left[ \pr(\Cqq^{1133}+\Cqqp^{3113},\Cqqp^{1133}+\Cqq^{3113})
+ \pr(\Cuu^{1133},\Cuu^{3113}) \right] \notag \\[1mm]
& & + \frac{\tilde D_1^{F,B}}{\Lambda^4}
\left[ \pr(\Cqup^{1331},\Cqu^{1331}) + \pr(\Cqup^{3113},\Cqu^{3113}) \right] \notag \\[1mm]
& & + \frac{0.0725~\text{pb}\cdot\text{TeV}^4}{\Lambda^4}
\pr(\Cqup^{3311},\Cqu^{3311}) \notag \\
& & - \frac{0.0395~\text{pb}\cdot\text{TeV}^4}{\Lambda^4}
\left[ \pr(\Cqq^{1133}+\Cqqp^{3113},\Cqup^{1331},\Cqqp^{1133}+\Cqq^{3113},\Cqu^{1331}) \right. \notag \\[1mm]
&& \left. + \pr(\Cqup^{3113},\Cuu^{1133},\Cqu^{3113},\Cuu^{3113}) \right] \,,
\end{eqnarray}
being the numerical constants
\begin{align}
& D_\text{int}^F = \tilde D_\text{int}^B = 0.499 ~\text{pb} \cdot \text{TeV}^2 \,,
&& D_\text{int}^B = \tilde D_\text{int}^F = 0.219 ~\text{pb} \cdot \text{TeV}^2 \,, \notag \\
& D_1^F = \tilde D_1^B = 0.0890 ~\text{pb} \cdot \text{TeV}^4 \,,
&& D_1^B = \tilde D_1^F = 0.0209 ~\text{pb} \cdot \text{TeV}^4 \,.
\end{align}
Obviously, $\sigma = \sigma^F + \sigma^B$.
The numerical coefficients of quadratic terms are about $1/5$ of the linear ones, so these terms can be ignored in a first approximation, provided that $\Lambda \gtrsim 1$ TeV and the operator coefficients are of order unity. For the latter process the forward and backward cross sections are
\begin{eqnarray}
\sigma^{F,B}(d \bar d) & = & 0.434~\text{pb}
+ \frac{D_\text{int}^{F,B}}{\Lambda^2} \left[ \Cqqp^{1133}+2\,\Cudp^{3311} \right]
- \frac{\tilde D_\text{int}^{F,B}}{\Lambda^2} \left[ \Cqu^{1331}+\Cqd^{3113}  \right] \notag \\
& & + \frac{D_1^{F,B}}{\Lambda^4}
\left[ \pr(\Cqq^{1133},\Cqqp^{1133}) + 4 \pr(\Cud^{3311},\Cudp^{3311}) \right] \notag \\[1mm]
& & + \frac{\tilde D_1^{F,B}}{\Lambda^4}
\left[ \pr(\Cqup^{1331},\Cqu^{1331}) + \pr(\Cqdp^{3113},\Cqd^{3113}) + \oh \pr(\Cqqep^{1331},\Cqqe^{1331}) \right] \notag \\
& & + \frac{9.86~\text{fb}\cdot\text{TeV}^4}{\Lambda^4}
\, \pr(\Cqqe^{3311},\Cqqep^{3311})
+ \frac{D_2^{F,B}}{\Lambda^4} \RE \pr(\Cqqe^{3311},\Cqqep^{1331},\Cqqep^{3311},\Cqqe^{1331})
 \notag \\[1mm]
& & - \frac{6.47~\text{fb}\cdot\text{TeV}^4}{\Lambda^4}
\left[\pr(\Cqq^{1133},\Cqup^{1331},\Cqqp^{1133},\Cqu^{1331}) \right. \notag \\[1mm]
& & \left. + 2 \pr(\Cqdp^{3113},\Cud^{3311},\Cqd^{3113},\Cudp^{3311})
\right] \,,
\end{eqnarray}
with the numerical constants
\begin{align}
& D_\text{int}^F = \tilde D_\text{int}^B = 0.0808 ~\text{pb} \cdot \text{TeV}^2 \,,
&& D_\text{int}^B = \tilde D_\text{int}^F = 0.0388 ~\text{pb} \cdot \text{TeV}^2 \,, \notag \\
& D_1^F = \tilde D_1^B = 12.1 ~\text{fb} \cdot \text{TeV}^4 \,,
&& D_1^B = \tilde D_1^F = 3.22 ~\text{fb} \cdot \text{TeV}^4 \,, \notag \\
& D_2^F = 5.42 ~\text{fb} \cdot \text{TeV}^4 \,,
&& D_2^B = 14.3 ~\text{fb} \cdot \text{TeV}^4 \,.
\end{align}
In this case the quadratic terms are multiplied by small numerical factors, and can be dropped in a first approximation.
The FB asymmetry can be obtained just summing the $u \bar u$ and $d\bar d$ contributions and using Eq.~(\ref{ec:afb}). Besides, we note that, among the seven operators which interfere with the SM amplitudes and could give sizeable contributions to the FB asymmetry, one ($O_{qq}^{3113}$) is also involved in single top production (see section~\ref{sec:5.1}). If the FB asymmetry can be explained by new physics which manifests as four-fermion effective operators, related effects might be seen in single top production at LHC as well.

\subsection{Like-sign top pair production in $pp$, $p \bar p$ collisions}
\label{sec:6.2}

The process $u_i u_j \to t t$ is rather interesting due to its potentially large cross section, in particular for the case of two initial $u$ valence quarks, and its striking signature of two like-sign top quarks.
For $i \neq j$ the matrix element is similar to the one for $t \to u_i u_i \bar u_j$, $u_i u_i \to t u_j$, except for some extra interference terms proportional to $m_t^2$, which did not appear in the former processes because $u_j$ was taken massless. The cross sections can be written as
\begin{eqnarray}
\sigma(uc,\bar u \bar c) & = & \frac{E_1}{\Lambda^4} \left[ |\Cqq^{1323}+\Cqqp^{1323}|^2 + |\Cuu^{1323}|^2 \right] \notag \\
& & + \frac{E_2}{\Lambda^4} \left[ \pr(\Cqup^{1323},\Cqu^{1323}) + \pr(\Cqup^{2313},\Cqu^{2313}) \right] \notag \\
& & + \frac{E_3}{\Lambda^4} \left\{ \RE \! \left[ \Cqup^{1323} \Cqu^{1323*} + \Cqup^{2313} \Cqu^{2313*} \right] \right. \notag \\
& & \left. + \osx \left[ |\Cqup^{1323}|^2 + |\Cqu^{1323}|^2 + |\Cqup^{2313}|^2
+ |\Cqu^{2313}|^2 \right] \right\} \,,
\end{eqnarray}
with $E_{1-3}$ numerical factors, whose values for LHC at 14 TeV, LHC at 7 TeV and Tevatron are given in Tables~\ref{tab:cross-like14}, \ref{tab:cross-like7} and \ref{tab:cross-like2}, respectively, using CTEQ6L1 PDFs with a factorisation scale $Q=m_t$.
The case $i=j$, with identical particles (except for colour) in both the initial and final states, is quite singular. One of the subtleties particular to these processes is that for different initial quark colours $a$, $b$ the gauge-invariant operators have two terms which contribute to the amplitude. For example, for $i=1$ we have
\begin{eqnarray}
(\bar u_R \gM t_R) (\bar u_R \gm t_R) & \to & (\bar u_{Ra} \gM t_{Ra}) (\bar u_{Rb} \gm t_{Rb}) + (\bar u_{Rb} \gM t_{Rb}) (\bar u_{Ra} \gm t_{Ra}) \notag \\
& & = 2 (\bar u_{Ra} \gM t_{Ra}) (\bar u_{Rb} \gm t_{Rb}) \,, \notag \\
(\bar u_L \gM t_L) (\bar u_R \gm t_R) & \to & (\bar u_{La} \gM t_{La}) (\bar u_{Rb} \gm t_{Rb}) + (\bar u_{Lb} \gM t_{Lb}) (\bar u_{Ra} \gm t_{Ra}) 
\end{eqnarray}
(no sum over $a,b$).
For equal colours, $u_a u_a \to t_a t_a$ there is only one such term but amplitudes get four contributions with a $1/2$ symmetry factor for identical particles, as usual.\footnote{Two contractions were missing from the amplitudes for $u u,c c \to t t$ in the first two versions of this paper. The cross section expressions have already been corrected in Ref.~\cite{AguilarSaavedra:2011zy}, and numerical values for LHC with a CM energy of 7 TeV have been given. By introducing a $t$-channel propagator in the amplitudes, the results have also been compared with the cross sections for a flavour-violating $Z'$ boson in Refs.~\cite{Bhattacherjee:2011nr,Berger:2011ua}, obtaining a good agreement.}
After colour averaging and phase space integration, the cross sections read
\begin{eqnarray}
\sigma(u_i u_i,\bar u_i \bar u_i) & = & \frac{E_1}{\Lambda^4} \left[ |\Cqq^{i3i3}+\Cqqp^{i3i3}|^2 + |\Cuu^{i3i3}|^2 \right] \notag \\
& & + \frac{E_2}{\Lambda^4} \left[ |\Cqup^{i3i3}|^2 + |\Cqu^{i3i3}|^2 + \twt \, \RE \Cqup^{i3i3} \Cqu^{i3i3*} \right] \notag \\
& & + \frac{E_3}{\Lambda^4} \left\{ \RE \Cqup^{i3i3} \Cqu^{i3i3*}
+ \os \left[ |\Cqup^{i3i3}|^2 + |\Cqu^{i3i3}|^2 \right] \right\} \,.
\end{eqnarray}
with the factors $E_{1-3}$ collected in Tables~\ref{tab:cross-like14}, \ref{tab:cross-like7} and \ref{tab:cross-like2}. The large numerical value of the coefficient $E_1$ for  initial $uu$ states at LHC, already with a CM energy of 7 TeV, implies an excellent sensitivity to the four-fermion operators $O_{qq^{(')}}^{1313}$ and
$O_{uu}^{1313}$, namely four-fermion terms $(\bar u_L \gM t_L) (\bar u_L \gM t_L)$
and $(\bar u_R \gM t_R) (\bar u_R \gM t_R)$. It is then expected that a large scale $\Lambda$ will be probed at LHC for these operators, in the clean final state of two like-sign top quarks.

\begin{table}[htb]
\begin{small}
\begin{center}
\begin{tabular}{lccccc}
$u_i u_j \to t t$  & $E_1$ & $E_2$  & $E_3$ \\
\hline
$i=1$ \quad $j=1$  & 75.6  & 9.60   & -0.930   & $t$ \\
                   & 0.859 & 0.114 & -0.0423  & $\bar t$ \\
$i=1$ \quad $j=2$  & 2.15  & 0.563  & -0.158   & $t$ \\
                   & 0.184 & 0.0496 & -0.0211  & $\bar t$ \\
$i=2$ \quad $j=2$ & 0.151 & 0.0205 & -0.00977 & $t/\bar t$
\end{tabular}
\end{center}
\end{small}
\caption{Numerical factors for like-sign top pair cross sections at LHC with 14 TeV. The units of $E_{1-3}$ are $\text{pb} \cdot \text{TeV}^4$. The labels $t$, $\bar t$ indicate whether the factors correspond to the processes in the left column or the charge conjugate.}
\label{tab:cross-like14}
\end{table}

\begin{table}[htb]
\begin{small}
\begin{center}
\begin{tabular}{lccccc}
$u_i u_j \to t t$  & $E_1$ & $E_2$  & $E_3$ \\
\hline
$i=1$ \quad $j=1$  & 15.8  & 2.05    & -0.420   & $t$ \\
                   & 0.102 & 0.0141  & -0.00783 & $\bar t$ \\
$i=1$ \quad $j=2$  & 0.314  & 0.0848  & -0.0389  & $t$ \\
                   & 0.0191 & 0.00534 & -0.00330 & $\bar t$ \\
$i=2$ \quad $j=2$ & 0.0138 & 0.00193 & -0.00132 & $t/\bar t$
\end{tabular}
\end{center}
\end{small}
\caption{Numerical factors for like-sign top pair cross sections at LHC with 7 TeV. The units of $E_{1-3}$ are $\text{pb} \cdot \text{TeV}^4$. The labels $t$, $\bar t$ indicate whether the factors correspond to the processes in the left column or the charge conjugate.}
\label{tab:cross-like7}
\end{table}

\begin{table}[htb]
\begin{small}
\begin{center}
\begin{tabular}{lccccc}
$u_i u_j \to t t$  & $E_1$ & $E_2$  & $E_3$ \\
\hline
$i=1$ \quad $j=1$ & 13.8   & 2.04    & -1.88    & $t/\bar t$ \\
$i=1$ \quad $j=2$ & 0.983  & 0.296   & -0.294   & $t/\bar t$ \\
$i=2$ \quad $j=2$ & 0.0204 & 0.00319 & -0.00380 & $t/\bar t$
\end{tabular}
\end{center}
\end{small}
\caption{Numerical factors for like-sign top pair cross sections at Tevatron. The units of $E_{1-3}$ are $\text{fb} \cdot \text{TeV}^4$. The labels $t/\bar t$ indicate that the factors are equal for the processes in the left column and the charge conjugate.}
\label{tab:cross-like2}
\end{table}

\subsection{Top pair production in $e^+ e^-$ collisions}
\label{sec:6.3}

We finally present results for $t \bar t$ production at a future ILC, including all four-fermion contributions and considering longitudinally polarised beams.
The SM cross sections for $e^+ e^- \to t \bar t$ are
\begin{eqnarray}
\sigma_\text{SM}(e_R^+ e_L^-) & = & \frac{\beta}{16\pi} \left\{
s(3+\beta^2) \left[ |\vll|^2 +|\vlr|^2 \right] + 24 \, m_t^2 \vll \vlr \right\} \,, \notag \\
\sigma_\text{SM}(e_L^+ e_R^-) & = & \frac{\beta}{16\pi} \left\{
s(3+\beta^2) \left[ |\vrl|^2 +|\vrr|^2 \right] + 24 \, m_t^2 \vrl \vrr \right\} \,, \notag \\[1mm]
\sigma_\text{SM}(e_L^+ e_L^-) & = & \sigma_\text{SM}(e_R^+ e_R^-) = 0 \,,
\label{ec:eettSM}
\end{eqnarray}
with $\beta = 1-4 m_t^2/s$ the top velocity in the CM frame. The ``effective'' couplings apprearing in these equations are
\begin{equation}
\mathcal{V}_{ij} = e^2 \left[ \frac{a_i^e a_j^t}{s_W^2 c_W^2 (s-M_Z^2)} - \frac{Q_t}{s} \right] \,, \quad i,j=L,R \,,
\end{equation}
being $Q_t=2/3$ the top quark charge, $s_W^2$ the electroweak mixing angle and
\begin{align}
& a_L^e = -\oh + s_W^2 \,, && a_R^e = s_W^2 \,, \notag \\
& a_L^t = \oh - \twt s_W^2 \,, && a_R^t = - \twt s_W^2
\end{align}
the chiral couplings of the electron and top quark to the $Z$ boson. Four-fermion vector terms can be included in Eqs.~(\ref{ec:eettSM}) simply by replacing 
\begin{align}
& \vll \to \vll + 2 \, \RE \alq^{1133} \,,
  && \vlr \to \vll - \RE \alu^{1331} \,, \notag \\
& \vrl \to \vrl - \RE \aqe^{3113} \,,
  && \vrr \to \vrr + 2 \, \RE \aeu^{1133} \,.
\end{align}
We find it more convenient, however, to give separately the interference of four-fermion operators with the SM and full quadratic four-fermion cross sections, including operators which do not interfere. The former is
\begin{eqnarray}
\sigma_\text{int}(e_R^+ e_L^-) & = & \frac{\beta}{8\pi \Lambda^2} \left\{
s(3+\beta^2) \left[ 2 \vll \,\RE \Clq^{1133} - \vlr \,\RE \Clu^{1331} \right] \right. \notag \\[1mm]
& & \left. + 12 \, m_t^2 \left[2 \vlr \,\RE \Clq^{1133}-\vll \,\RE \Clu^{1331} \right]
 \right\} \,, \notag \\
\sigma_\text{int}(e_L^+ e_R^-) & = & \frac{\beta}{8\pi\Lambda^2} \left\{
s(3+\beta^2) \left[ 2 \vrr \, \RE \Ceu^{1133} - \vrl \, \RE \Cqe^{3113} \right] \right. \notag \\[1mm]
& & \left. + 12 \, m_t^2 \left[ 2 \vrl \, \RE \Ceu^{1133} - \vrr \, \RE \Cqe^{3113}
 \right] \right\} \,,
\end{eqnarray}
with obviously $\sigma_\text{int}(e_L^+ e_L^-) = \sigma_\text{int}(e_R^+ e_R^-) = 0$.
The four-fermion polarised cross sections are
\begin{eqnarray}
\sigma_\text{4F}(e_R^+ e_L^-) & = & \frac{\beta}{16\pi\Lambda^4} \left\{
s(3+\beta^2) \left[ 4(\RE \Clq^{1133})^2 +(\RE \Clu^{1331})^2 \right] \right. \notag \\[1mm]
& & \left. - 48 \, m_t^2 \, \RE \Clq^{1133} \, \RE \Clu^{1331} \right\} \,, \notag \\
\sigma_\text{4F}(e_L^+ e_R^-) & = & \frac{\beta}{16\pi\Lambda^4} \left\{
s(3+\beta^2) \left[ 4(\RE \Ceu^{1133})^2 + (\RE \Cqe^{3113})^2\right] \right. \notag \\[1mm]
& & \left. - 24 \, m_t^2 \, \RE \Ceu^{1133} \, \RE \Cqe^{3113} \right\} \,, \notag \\
\sigma_\text{4F}(e_L^+ e_L^-) & = & \sigma_\text{4F}(e_R^+ e_R^-) = \frac{\beta}{64\pi\Lambda^4} \left\{
s(3+\beta^2) |\Cqle^{3113}|^2 \right. \notag \\[1mm]
& & \left. + 6s(1+\beta^2) \left[ |\Clqe^{1133}|^2
+\RE \Cqle^{3113} \Clqe^{1133*} \right] \right\} \,.
\end{eqnarray}
Our expressions agree with the unpolarised cross sections in Ref.~\cite{Grzadkowski:1995te}
for vector terms, as it can be seen with the redefinitions
$A_L \equiv 2 \,\RE \alq^{1133} -\RE \alu^{1331}$,
$B_L \equiv 2 \,\RE \alq^{1133} + \RE \alu^{1331}$,
$A_R \equiv 2 \,\RE \aeu^{1133}-\RE \aqe^{3113}$,
$B_R \equiv -2 \,\RE \aeu^{1133}-\RE \aqe^{3113}$. For scalar terms our expressions agree as well, where comparable. Note also that cross sections for transverse beam polarisation have been recently given in Ref.~\cite{Ananthanarayan:2010xs}.

\section{Summary}
\label{sec:7}

In this paper we have thoroughly studied the role of gauge-invariant four-fermion operators in top physics. The first difficulty one has to address in such a study is merely to collect all the relevant four-fermion operators, which is cumbersome because of the large number of flavour combinations, not all of them independent, that can be written.

We have used a new, minimal four-fermion operator basis, which offers some advantages for calculations, to classify all gauge-invariant four-fermion operators giving terms with one or two top quarks. (Only a handful of operators give three or four top quarks, and their classification is straightforward.) We have given our results in several compact tables in which the Lagrangian terms can be directly read by intersecting the desired row and column. Having all the possible terms classified represents a good share of the work needed for any calculation, and so we expect that the tables provided will be useful for future studies. A bonus of this classification is that contributions from the same gauge-invariant operators to different channels can be easily related. Just as an example: we can identify which operators produce $t \bar b \bar u d$ terms (and thus contribute to single top production), which ones give $t \bar t \bar u u$ terms (contributing to top pair production) and those producing both.

We have gone beyond the classification, which is already important on its own, to provide calculations of all decay widths, single top and top pair production cross sections mediated by four-fermion operators, including the SM contribution when present. These calculations will be valuable to guide future more detailed simulations, not only to have ``reference'' numerical values to compare with, but also to identify the most useful channels and the relevant operators which can be probed. In this respect, we have found that in $s$-channel single top and $t \bar t$ production there are four-fermion operators which do not interfere with the SM amplitudes but whose quadratic $1/\Lambda^4$ contributions to the cross sections can be as large as the linear $1/\Lambda^2$ ones from the interfering ones. As we have argued in the introduction, quadratic corrections from such operators can and must be included in a complete analysis, even if we have ignored sub-leading effects from dimension-eight operators.

The phenomenological implications of our results have not been fully addressed, for example providing expected limits on operator coefficients. This work is left for future detailed studies. Still, there are several interesting points which are worth remarking here:
\begin{enumerate}
\item Four-fermion terms with fields $t \bar d_k \bar u_i d_j$ (including $d_k = b$) and $t \bar u_k \bar u_i u_j$ will be better probed in single top production than in top decays. In the former case, they contribute to SM $t$- and $s$-channel production, while in the latter they mediate new, FCN processes.
\item Four-fermion terms $t \bar d_k e_i \bar \nu_j$ (including $d_k=b$) can only be probed in top decays. Moreover, some of the gauge-invariant operators producing these terms, for example $O_{\ell q'}$, cannot be investigated in other processes as
single top or top pair production in $e^+ e^-$ collisions. The net contribution to the decay width of four-fermion operators is very small, but the interference produces an asymmetry in the distribution for invariant masses $m_{e\nu} < M_W$ and  $m_{e\nu} > M_W$, which should be studied with more detail.
\item Operators giving $t t \bar u_i \bar u_j$ terms will likely be probed with a good precision in like-sign top pair production at LHC, especially for $i=j=1$, where the cross sections are potentially large and the final state relatively clean.
\end{enumerate}
We have also given expressions to calculate the FB asymmetry for $t \bar t$ production at Tevatron including all contributing four-fermion operators, to complement present studies~\cite{Jung:2009pi}.

In summary, we have provided in this paper a roadmap for future studies of gauge-invariant four-fermion operators in top physics, which we expect will be useful now that the era of precision measurements in the top sector has just begun.
\section*{Acknowledgements}

I thank F. del \'Aguila, J. de Blas, M. P\'erez-Victoria, and J. Santiago for interesting discussions.
This work has been partially supported by a MICINN Ram\'on y Cajal contract and by projects FPA2006-05294 (MICINN), FQM 101 and FQM 437 (Junta de Andaluc\'{\i}a),
CERN/FP/83588/2008 (FCT),
and by the European Community's Marie-Curie Research Training
Network under contract MRTN-CT-2006-035505 ``Tools and Precision
Calculations for Physics Discoveries at Colliders''.

\end{document}